\documentclass[10pt]{iopart}
\usepackage{iopams}  
\usepackage{graphicx}
\usepackage{epsfig}
\usepackage{epstopdf}
\newcommand{\nn}{\nonumber\\}

\newcommand{\vep}{\varepsilon}
\newcommand{\del}{\Delta}
\newcommand{\be}{\begin{equation}}
\newcommand{\bse}{\begin{subequation}}
\newcommand{\ese}{\end{subequation}}
\newcommand{\ee}{\end{equation}}
\newcommand{\eei}{\end{eqnarray}\indent\indent}
\newcommand{\bc}{\begin{center}}
\newcommand{\ec}{\end{center}}
\newcommand{\ba}{\begin{array}}
\newcommand{\ea}{\end{array}}

\newcommand{\hs}{\,-\,}
\newcommand{\p}{\partial}

\newcommand{\car}{{\cal R}}

\newcommand{\sfrac}[2]{{\textstyle{#1\over#2}}}
\def\case#1/#2{\textstyle\frac{#1}{#2} }
\newcommand{\nb}{\nabla}

\newcommand{\gd}{g_{ab}}
\newcommand{\tl}{\tilde}
\newcommand{\bver}{\begin{verbatim}}
\newcommand{\ever}{\end{verbatim}}

\newcommand{\tlnb}{\tilde{\nabla}^{2}}
\newcommand{\np}{\newpage}
\def\cqg{{\em Class. Quantum Grav.\/} }

\begin{document}
%%%%%%%%%%%%%%%%%%%%%%%%%%%%%%%%%%%%%%%%%%%%%%%%%%%%%
\title{Covariant gauge-invariant perturbations in multifluid $f(R)$ gravity}
%%%%%%%%%%%%%%%%%%%%%%%%%%%%%%%%%%%%%%%%%%%%%%%%%%%%%
\author{Amare Abebe $^{1,2}\footnote{amare.abebe@acgc.uct.ac.za}$, Mohamed Abdelwahab $^{1,2}$ , \'Alvaro\,de la Cruz-Dombriz$^{1,2}$ and Peter K. S. Dunsby$^{1,2,3}$}
\address{$^{1}$  Astrophysics, Cosmology and Gravity Centre (ACGC), University of Cape Town, Rondebosch, 7701, South Africa}
\address{$^{2}$ Department of Mathematics and Applied Mathematics, University of Cape Town, 7701 Rondebosch, Cape Town, South Africa}
\address{$^{3}$  South African Astronomical Observatory,  Observatory 7925, Cape Town, South Africa.}
\date{\today}
\begin{abstract}
We study the evolution of scalar cosmological perturbations in the $1+3$ covariant gauge-invariant formalism for generic $f(R)$ theories of gravity.  Extending previous works, we give a complete set of equations describing the evolution of matter and curvature fluctuations for a multi-fluid cosmological medium. We then specialize to a radiation-dust fluid described by barotropic equations of state and solve the perturbation equations around a background solution of $R^{n}$ gravity.  In particular we study exact solutions for scales much smaller and much larger than the Hubble radius and show that $n>\frac{2}{3}$ in order to have a growth rate compatible with the M\'esz\'aros effect.
\end{abstract}
\maketitle
%%%%%%%%%%%%%%%%%%%%%%%%%%%%%%%%%%%%%%%%%%%%%%%%%%%%%
\section{Introduction} 
%%%%%%%%%%%%%%%%%%%%%%%%%%%%%%%%%%%%%%%%%%%%%%%%%%%%%
%
There are at least three areas where standard General Relativity (GR) theory faces serious challenges from other competing fundamental theories \cite{moffat}. Firstly, attempts to unite quantum mechanics with GR have so far been unsuccessful and with the remarkable success the former has achieved over the last hundred years, many are questioning the unique status of GR,  suggesting that a fundamental modification of the theory could lead to a complete description of gravity on all scales.  Secondly,  in standard particle physics all the forces of nature save gravity have been shown to be  manifestations of an underlying  unified theory of nature and there are hopes that modifications of GR could lead to a grand unification of all the known forces into a single theory of everything. Thirdly and  most important to us here is the missing link between the observed universe and the standard theory of cosmology.

The  recent discovery of the accelerated expansion of the Universe has cast a new shadow on the simplest cosmology based on Einstein's 
theory of GR, together with a conventional matter description. Observational analyses \cite{Spergel}-\cite{bennett} show that only a tiny fraction ($\sim 4\% $) of the energy content of the Universe is known to exist in normal matter form, whereas $\sim 23\%$ of it exists in a little understood {\it dark matter} form. Dubbed as {\it Dark Energy} (DE),  the remaining ($\sim 72\%$) of the energy content of the Universe is believed to be the cause of the inferred accelerated cosmic expansion.

Many candidates have been put forward as an explanation for DE, but most of them fall under one of these three forms: the cosmological constant, exotic scalar fields (such as Quintessence) and geometrical dark energy in which the gravitational Lagrangian is modified with respect to the usual Einstein-Hilbert one (see  \cite{Ref1}, \cite{Ref2}, \cite{Ref3} and \cite{Ref4} for an extensive review). 

An important class of modified gravity are the scalar-tensor and $f(R)$ theories \cite{Ref5}-\cite{Ref15}.  Both these candidates have their own
serious shortcomings \cite{Ref11, weinberg, jack},  and have to pass rigorous theoretical and observational scrutiny before they can be accepted as viable theories \cite{Silvestri, Capozziello:2011et}.  In this work we will only concentrate on the interpretation of DE as geometrical manifestation of a more fundamental theory, focusing on $f(R)$-gravity.

It is well known that the dynamical evolution of small density perturbations, seeded in the early Universe, led to the large-scale structure we see today \cite{JMB}-\cite{CMB}. An excellent framework for studying cosmological perturbations is the 1+3 covariant approach, which has been developed, among other things,  to analyze the evolution of linear perturbations of  Friedmann-Lema\^{i}tre-Robertson-Walker (FLRW) models in GR  \cite{EB} - \cite{PPhD}. 

In recent years, higher order theories of gravity have attracted a great deal of attention. A detailed analysis of the background FLRW models using dynamical system techniques has shown that there exist classes of fourth order theories which admit a transient decelerated expansion phase during which structure formation can take place, followed by a DE-like era which drives the present cosmological acceleration (see \cite{Ref16} among many others). 
However, it has been
proved \cite{Dombriz_PRD} that when dust matter scalar cosmological perturbations
are studied in the metric formalism, $f(R)$ theories, even mimicking the standard
cosmological expansion, provide a different matter power spectrum from 
that predicted by the $\Lambda{\rm CDM}$ model \cite{Dombriz_PRL}.
In \cite{CDT} the evolution of scalar perturbations of FLRW models in fourth order gravity was developed for single barotropic fluids using the 1+3 covariant approach and the solutions of the perturbation equations on large-scales showed that a decelerated phase is not necessarily required to form large scale structures. This divergence from the standard GR provides us with a distinguishable signature of fourth order theories, which can be tested against observations.

However, since the Universe consists of a mixture of  fluids, a complete treatment of perturbations in fourth order theories requires taking this into account. The aim of this paper is therefore to present a general framework for studying multi-fluid cosmological perturbations with a 
completely general equation of state in a $f(R)$ theory of gravity, using the 1+3 covariant approach.

This paper has been organized as follows: in section 2 we give the general outline of the 1+3 covariant approach. In section 3 we discuss the choice of frame and define the key variables used in the description of perturbations in the total fluid and the individual fluid components. Equations for these variables are given in section 4. In  sections 5 and 6, respectively,  the scalar and the harmonically decomposed forms of our equations are presented. Applications to a radiation-dust cosmological medium are given in section 7, with sections 8 and 9 devoted to the analysis of the short and long wavelength limits of the perturbation equations. We conclude the paper by giving our conclusions in section 10.
 
The natural unit conventions $(\hbar  = c = k_{B} = 8\pi G = 1)$  are in use. Latin indices of tensors run from $0$ to $3$. The symbols $\nb$ and $;$ represent the usual covariant derivative,  $\p$ corresponds to partial differentiation and an over-dot shows differentiation with respect to proper time. We use the ($-+++ $)  spacetime signature in this work.

 For an arbitrary $f(R)$ gravity the generalized Einstein-Hilbert  action can be written as
 \be
 {\cal{A}}_{f(R)}=\frac{1}{2}\int{{\rm d}^{4}x\sqrt{-g}\left(f(R)+2{\cal{L}}_{m}  \right) }\;.
 \ee
 A generalization of the Einstein's Field Equations (EFEs)  derived by varying this action with respect to the metric takes the form
 \be \label{efe1}
 f'G_{ab}=T^{m}_{ab}+\frac{1}{2}g_{ab}(f-Rf')+\nb_{b}\nb_{a} f'-\gd \nb_{c}\nb^{c} f'\;,
 \ee
 where  $f\equiv f(R)$, $f'\equiv \frac{{d}f}{{d}R}$ and
$T^{m}_{ab}\equiv \frac{2}{\sqrt{-g}}\frac{\delta(\sqrt{-g}\mathcal{L}_{m})}{\delta \gd}$.

Defining the energy momentum tensor of the curvature ``fluid'' as
\be
T^{R}_{ab}\equiv\frac{1}{f'}\left[\frac{1}{2}(f-Rf')g_{ab}+\nb_{b}\nb_{a}f'-g_{ab}\nb_{c}\nb^{c}f' \right],
\ee
the field equations (\ref{efe1}) can be written in a more compact form
 \be
 G_{ab}=\tl T^{m}_{ab}+T^{R}_{ab}
 \equiv T_{ab},
 \ee
 where the effective energy momentum tensor of standard matter is given by
 \be
 \tl T^{m}_{ab}\equiv\frac{T^{m}_{ab}}{f'}\;.
 \ee
  
 Assuming that the energy-momentum conservation of standard matter $T^{m}_{ab}{}^{;b}=0$ holds,  leads us to conclude that $T_{ab}$ is divergence-free, i.e., $T_{ab}{}^{;b}=0$, and therefore  $\tl{T}^{m}_{ab}$ and $T^{R}_{ab}$ are not individually conserved \cite{CDT}:
 \be
 \tl T^{m}_{ab}{}^{;b}=\frac{T^{m}_{ab}{}^{;b}}{f'}-\frac{f''}{f^{'2}}T^{m}_{ab}R^{;b}\;,~~~T^{R}_{ab}{}^{;b}=\frac{f''}{f^{'2}}T^{m}_{ab}R^{;b}\;.
 \ee
%%%%%%%%%%%%%%%%%%%%%%%%%%%%%%%%%%%%%%%%%%%%%%%
\section{Covariant Decomposition of $4^{th}$-order Gravity}
%%%%%%%%%%%%%%%%%%%%%%%%%%%%%%%%%%%%%%%%%%%%%%%
\subsection{Preliminaries}
%%%%%%%%%%%%%%%%%%%%%%%%%%%%%%%%%%%%%%%%%%%%%%%%
The standard perturbation theory based on the metric formalism has disadvantages when it comes to extracting physical information from the perturbation variables. For example, it requires a complete specification of the correspondence between the lumpy,  perturbed universe and the background spacetime. In other words, this approach is \textit{gauge-dependent}. In this paper we instead use the 1+3-covariant formalism, a fluid approach, which, when applied to cosmological perturbations leaves no unphysical modes in the evolution of the fluctuations: it requires no prior metric specification and is \textit{gauge-invariant} by construction.
%%%%%%%%%%%%%%%%%%%%%%%%%%%%%%%%%%%%%%%%%%%%%%%%%
\subsection{Kinematics}
%%%%%%%%%%%%%%%%%%%%%%%%%%%%%%%%%%%%%%%%%%%%%%%%%
We project onto surfaces orthogonal to the $4$-velocity of the fluid flow using the projection tensor $h_{ab}\equiv g_{ab}+u_{a}u_{b}$ and $\tl\nb_{a}=h^{b}_{a}\nb_{b}$ is the spatially totally projected covariant derivative operator orthogonal to $u^{a}$. The covariant convective and spatial covariant  derivatives on a scalar function $X$ are respectively given by
\be
\dot{X}=u_{a}\nb^{a}X, ~~~~~\tl \nb_{a}X=h_{a}{}^{b}\nb_{b}X\;.
\ee
The geometry of the flow lines is  determined by  the kinematics of $u^{a}$:
\begin{eqnarray}
&\nb_{b}u_{a}=\tl \nb_{b}u_{b}-a_{a}u_{b}\;, \label{delua}\\ 
&\tl \nb_{b}u_{a}=\sfrac{1}{3}\Theta h_{ab}+\sigma_{ab}+\omega_{ab}\;.
\label{projdelua}
\end{eqnarray}
From (\ref{delua}) and (\ref{projdelua}) we obtain an  important equation relating our key kinematic quantities:
\be
\nb_{b}u_{a}=-u_{b}\dot{u}_{a}+\sfrac{1}{3}\Theta h_{ab}+\sigma_{ab}+\omega_{ab}\;,
\ee
The RHS of this equation contains the acceleration of the fluid flow $\dot{u}_{a}$, expansion $\Theta$, shear $\sigma_{ba}$ and vorticity $\omega_{ba}$.

Another key equation is the propagation equation for the expansion -  the Raychaudhuri  equation (given here for the FLRW background) :
\be
\dot{\Theta}+\frac{1}{3}\Theta^{2}+\frac{1}{2}(\mu^{}+3p^{})=0\;,
\ee 
where $\mu$ and $p$ hold for the total energy density and isotropic pressure respectively. This equation together with the equation of state $p=p(\mu,s)$, the energy conservation equation
\be
\dot{\mu}+\Theta(\mu+p)=0\;,
\ee
 and the Friedmann equation
 \be
 \Theta^{2}+\frac{9K}{a^{2}}-3\mu=0\;,
 \ee
form a closed system of equations and completely characterize the kinematics of the background cosmological model.
 
In this paper angular brackets denote the projection of a tensorial quantity onto the tangent 3-space. Thus the relations
\begin{eqnarray}\label{vector}
&V_{\langle a\rangle}=h_{a}{}^{b}{}V_{b}\;,\\ \label{tensor}
&W_{\langle ab\rangle}=\left[h_{(a}{}^{c}{}h_{b)}{}^{d}-\sfrac{1}{3}h^{cd}h_{ab} \right]W_{cd} 
\end{eqnarray}
give the projection of a vector $V_{a}$ and the projected, trace-free part of a tensor $W_{ab}$ respectively.
%%%%%%%%%%%%%%%%%%%%%%%%%%%%%%%%%%%%%%%%%%%%%%%%%%
\section{Matter Description}\label{sec: variables}
%%%%%%%%%%%%%%%%%%%%%%%%%%%%%%%%%%%%%%%%%%%%%%%%%%
\subsection{Effective Total Energy-Momentum Tensor}
%%%%%%%%%%%%%%%%%%%%%%%%%%%%%%%%%%%%%%%%%%%%%%%%%%
The thermodynamical description of a relativistic fluid is dictated by the energy momentum tensor $T_{ab}$, the particle flux $N^{a}$ and the entropy flux $S^{a}$ of the system. Whereas $T^{ab}$ and $S^{a}$ always satisfy respectively the conservation of 4-momentum and the second law of thermodynamics, namely  
\be
T^{ab}{}{}_{;b}=0\;, ~~~S^{a}{}_{;a} \geq 0\;,
\ee
particle flux conservation, i.e.,   $N^{a}{}_{;a}=0$, may be violated.

The total energy-momentum tensor in a general frame is sourced by $\mu$, $p$, the energy flux $q_{\langle a\rangle}$, and the 
anisotropic pressure $\pi_{\langle ab\rangle}$:
\be \label{EMT}
T_{ab}=\mu u_{a}u_{b}+ph_{ab}+2q_{(a}u_{b)}+\pi_{ab}=\tl{T}^{m}_{ab}+T^{R}_{ab}\;.
\ee
It  defines our thermodynamical quantities:
\begin{eqnarray}
&\mu^{tot}=T^{tot}_{ab}u^{a}u^{b}=\tl \mu_{m}+\mu_{R}\;,\\
&p^{tot}=\frac{1}{3}T^{tot}_{ab}h^{ab}=\tl p_{m}+p_{R}\;,\\
&q^{tot}_{a}=-T^{tot}_{bc}h^{b}{}_{a}u^{c}=\tl q^{m}_{a}+q^{R}_{a},\\
&\pi^{tot}_{ab}=T^{tot}_{cd}h^{c}_{\langle a}h^{d}_{b\rangle}=\tl \pi^{m}_{ab}+\pi^{R}_{ab}\;,
\end{eqnarray}
where $\tl \mu_{m}=\frac{\mu_{m}}{f'}, ~~~\tl p_{m}=\frac{p_{m}}{f'}, ~~~\tl q^{m}_{a}=\frac{q^{m}_{a}}{f'},~ \mbox{and~~}\tl \pi^{m}_{ab}=\frac{\pi^{m}_{ab}}{f'}$ are the effective  thermodynamic quantities of matter.

 If  we impose the  Strong Energy Condition $T_{ab}V^{a}V^{b} \geq 0$ for all timelike vectors $V^{a}$, then
  $ T_{ab}$ will have a  unique unit timelike vector $u^{a}_{E}$ $(u^{a}_{E}u^{E}_{a}=-1)$. Another timelike vector
$u^{a}_{N}$ can be  defined along the flux $N^{a}_{N}$, i.e., $u^{a}_{N}=\frac{N^{a}}{\sqrt{-N{b}N^{b}}}$.

 For a perfect fluid (or an unperturbed fluid in the background space), $u^{a}_{E}$, $u^{a}_{N} $ and $S^{a}$ are all parallel \cite{DBE} and a unique hydrodynamic 4-velocity $u^{a}$ can be defined for the fluid flow, in which case
\be\label{perfectEMT}
T_{ab}=\mu u_{a}u_{b}+ph_{ab}\;,~~~
N^{a}=nu^{a},~~~S^{a}=su^{a}\;,
\ee
where $\mu$ and $p$ are related by the equation of state
$p=p(\mu,s)$.
$n=-N^{a}u_{a}$ and $s=-S_{a}u^{a}$ define the particle and entropy densities respectively in the local rest frame of an observer attached to $u^{a}$.

We can also decompose the EMT with respect to another frame, say $n^{a}$, but in this case we need to introduce a particle drift  $\tilde{j}^a=\tilde{h}^{a}{}_{b}N^{a}$ \cite{DBE,BDE,KE}.

If the fluid  is imperfect, the fluid hydrodynamic 4-velocity is  no longer unique and  our EMT will take the more general form given above Eqn. (\ref{EMT}) and the particle flux includes a \textit{drift} term:
\be
N^{a}=nu^{a}+j^{a}\;.
\ee
Choosing the relevant frame is a crucial step in the covariant formulation of perturbation theories, since $u^{a}$ is the velocity of fundamental observers in the Universe
\footnote{Fluid flow vector $u^{a}$ is uniquely defined as the future directed timelike eigenvector of the Ricci tensor:
$u^{a}=\frac{dx^{a}}{d\tau}$, where $x^{a}(\tau)$ describes the worldline of the fluid in terms of the proper time $\tau$. In our multi-fluid picture, it corresponds to the normal to the surface of homogeneity.}.

In the particle frame $u^{a}=u^{a}_{N}$, called the Eckart choice, an observer $O_{u=u_{N}}$ sees no particle drift and hence $j^{a}=j^{a}_{N}=0$. If, on the other hand, we consider the energy frame $u^{a}=u^{a}_{E}$,  also known as the Landau choice, an observer $O_{u_{E}}$ measures no energy flux ($q_{a}=q^{E}_{a}=0$) along the flow line and the EMT takes the form (\ref{perfectEMT}).

For multi\hs component  matter fluids we have 
\be
T^{m}_{ab}=\sum_{i}{T^{i}_{ab}}\;,
\ee
where
\begin{eqnarray}
&T^{i}_{ab}=\mu_{i}u^{i}_{a}u^{i}_{b}+p_{i}h^{i}_{ab}+q^{i}_{a}u^{i}_{b}+q^{i}_{b}u^{i}_{a}+\pi^{i}_{ab}\;,\\
&h^{i}_{ab}=g_{ab}+u^{i}_{a}u^{i}_{b}\;,\\
&N^{a}_{i}=n_{i}u^{a}_{i}+j^{a}_{i}\;,
\end{eqnarray}
 $u^{i}_{a}$  being the normalized fluid 4-velocity vector for the $i^{th}$ component, $u^{a}_{i}u^{i}_{a}=-1$,
which we can fix  by either choosing the energy frame $u^{a}_{i}=u^{a}_{Ei}$ thereby setting
$ q^{a}_{i}=q^{a}_{Ei}=0$,
or the particle frame $u^{a}_{i}=u^{a}_{Ni}$ for which
$j^{a}_{i}=j^{a}_{Ni}=0$ for that component. The velocity of the $i^{th}$ fluid component relative to the fundamental observer $O_{u}$ 
is defined to be  
 \be
 V^{a}_{i}\equiv u^{a}_{i}-u^{a}\;.
 \ee
 $V^{a}_{i}\neq 0$ for \textit{tilted}, inhomogeneous cosmological media whereas the special case where
$u^{a}_{i}$ coincides with $u^{a}$ describes an \textit{untilted} homogeneous cosmological medium.
Decomposition of the  matter stress-energy tensor with respect to the 4-velocity $u^{a}$ gives the following thermodynamical quantities:
\begin{eqnarray}
&\mu^{m}= T^{m}_{ab}u^{a}u^{b}=\sum^{N}_{i=1}{\mu_{i}}\;,\\
&p^{m}=\frac{1}{3}T^{m}_{ab}h^{ab}= \sum^{N}_{i=1}{ p_{i} }\;,\\
&q^{m}_{a}=-T^{m}_{bc}h^{b}_{a}u^{c}=\sum^{N}_{i=1}{(\mu_{i}+p_{i})V^{i}_{a}}\;,\\
&\pi^{m}_{ab}=T^{m}_{cd}h^{c}_{\langle a}h^{d}_{b\rangle}=0~~~\mbox{(to~first~order)}.
\end{eqnarray}
In a similar way we can decompose the  energy momentum tensor of the curvature fluid to obtain the corresponding  thermodynamical quantities (denoted in what follows by a ${\it{R}}$ superscript or subscript). All these quantities, unlike their matter counter-parts, vanish in standard GR, with a FLRW geometry: 
\begin{eqnarray}
&\mu^{R}=T^{R}_{\;ab}u^{a}u^{b}= \frac{1}{f'}\left[\frac{1}{2}(Rf'-f)-\Theta f'' \dot{R}+ f''\tilde{\nabla}^{2}R \right]\;,\\
&p^{R}=\frac{1}{3}T^{R}_{\;ab}h^{ab}=\frac{1}{f'}\left[\frac{1}{2}(f-Rf')+f''\ddot{R}+f'''\dot{R}^{2}\right.\nn
&\left.~~~~~~~~~~~~~~~~~+\frac{2}{3}\left( \Theta f''\dot{R}-f''\tilde{\nabla}^{2}R -f'''\tilde{\nabla}^{a}R \tilde{\nabla}_{a}R  \right)\right]\;,\\
&q^{R}_{a}=-T^{R}_{\;bc}h^{b}_{a}u^{c}=-\frac{1}{f'}\left[f'''\dot{R}\tilde{\nabla}_{a}R +f''\tilde{\nabla}_{a}\dot{R}-\frac{1}{3}f''\Theta \tilde{\nabla}_{a}R \right]\;,\\
&\pi^{R}_{ab}=T^{R}_{\;cd}h^{c}_{\langle a}h^{d}_{b\rangle}=\frac{1}{f'}\left[f''\tilde{\nabla}_{\langle a}\tilde{\nabla}_{b\rangle}R +f'''\tilde{\nabla}_{\langle a}R\tilde{\nabla}_{b\rangle}R-\sigma_{ab}\dot{R} \right]\;.
\end{eqnarray}
\begin{figure*}[h!]
	\centering
		\includegraphics[width=0.9\textwidth]{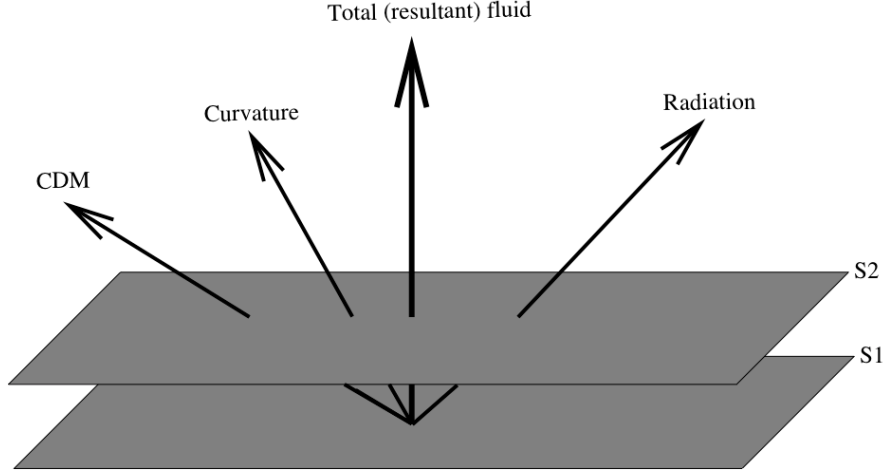}\,\,\,\,\,\,\,\,\,\,\,\,\,\,
		\caption{\footnotesize{The Multi-fluid diagram: The different arrows show the unit time-like four-velocity vectors at different hyper-surfaces S1 and S2. The vectors do not coincide at the perturbative level.	
		}}
	\label{fig:rangoespin}
\end{figure*}
In the background FLRW universe, $V^{a}_{i}=0$  and all perfect fluid components have the same $4$-velocity. By applying the  Stewart-Walker Lemma \cite{PPhD}, we can show that $V^{a}_{i}$ is a first-order gauge-invariant  (GI) quantity. If we choose the  fluid flow vector $u^{a}$ to coincide with the energy frame $u^{a}_{E}$ (see Fig.1 above), then exact FLRW models will be characterized by vanishing shear and vorticity of $u^{a}$ and all spatial gradients orthogonal to $u^{a}$ of any scalar quantity  \cite{BDE}:
\be
\sigma_{ab}=\omega_{ab}=0,~~~~~~\tl\nb_{a} X =0\;.
\ee
It then follows that, since 
\be
X_{a}=\tl\nb_{a}\mu=0,~~~Y_{a}=\tl\nb_{a}p,~~~Z_{a}=\tl\nb_{a}\Theta=0
\ee
in the background, then $\mu=\mu(t)$, $p=p(t)$ and $\Theta=\Theta(t)$. This necessitates the energy momentum tensor having the perfect fluid form, and hence the vanishing of the anisotropic pressure  $\pi_{ab}$ and the energy flux $q_{a}$.
%%%%%%%%%%%%%%%%%%%%%%%%%%%%%%%%%%%%%%%%%%%%%%%%%%%%%%%%%%
\subsection{Standard Inhomogeneity Variables for the Total Matter}
%%%%%%%%%%%%%%%%%%%%%%%%%%%%%%%%%%%%%%%%%%%%%%%%%%%%%%%%%%
The key variables characterizing the inhomogeneities of matter are 
\begin{eqnarray}
&D^{m}_{a}=a\frac{\tilde{\nabla}_{a}\mu_{m}}{\mu_{m}}\;,~~~~~~~~~~~~
Y_{a}=\tilde{\nabla}_{a}p_{m},\nn
&Z_{a}=a\tilde{\nabla}_{a}\Theta,~~~~~~~~~~~~~~~~C_{a}=a\tl \nb_{a}\tl R,\nn
&\varepsilon_{a}=\frac{a}{p_{m}}\left(\frac{\p p}{\p s}\right)\tilde{\nabla}_{a}s\;,~~~~~~~
A=a^{a}{}_{;a}=\tl\nb_{a}a^{a}\;,\nn
&A_{a}=\tilde{\nabla}_{a}A\;,~~~~~~~~~~~~~~~~~~
Q=q^{a}_{\,\,\,\,;a} \simeq \tilde{\nabla}_{a}q\;,
\end{eqnarray}
where $\it {a} \equiv a(t)$ here is the usual FLRW cosmological scale factor.
$D^{m}_{a}$ and $Z_{a}$ define the comoving fractional density gradient and comoving gradient of the expansion respectively and can in principle be measured observationally \cite{BDE}. The relation
\be\label{pvep}
p\varepsilon_{a}=\sum_{i}{p_{i}\varepsilon^{i}_{a}}+\frac{1}{2}\sum_{i,j}{\frac{h_{i}h_{j}}{h}(c^{2}_{si}-c^{2}_{sj})S^{ij}_{a}}
\ee
defines the dimensionless variable $\vep_{a}$ that quantifies entropy perturbations in the total fluid.  We have defined the shorthand $h\equiv \mu_{m}+p_{m}$ for the total matter fluid and $h_{i} \equiv \mu_{i}+p_{i}$ for the component matter fluids. $w$ and $c^{2}_{s}$ denote the effective barotropic equation of state and speed of sound of the total matter fluid, respectively, and are defined by
\be
w\equiv\frac{p_{m}}{\mu_{m}},~~~~~c^{2}_{s}\equiv\frac{\p p_{m}}{\p \mu_{m}}\;,\label{omegg}
\ee
whereas for each component mater fluid, these two quantities are given by 
$w_{i}\equiv p_{i}/\mu_{i}$ and $c^{2}_{si}\equiv\p p_{i}/\p \mu_{i}$.
%%%%%%%%%%%%%%%%%%%%%%%%%%%%%%%%%%%%%%%%%%%%%%%%%%%%%%%%%%
\subsection{Matter Inhomogeneity Variables for the Components}
%%%%%%%%%%%%%%%%%%%%%%%%%%%%%%%%%%%%%%%%%%%%%%%%%%%%%%%%%%
The variables characterizing inhomogeneities of matter for  the $i^{th}$- component fluid are defined as
\be
D^{i}_{a}=a\frac{\tilde{\nabla}_{a}\mu_{i}}{\mu_{i}}\;,~~~~~
Y^{i}_{a}=\tilde{\nabla}_{a}p_{i}\;,~~~~~
\varepsilon^{i}_{a}=\frac{a}{p^{i}}\left(\frac{\p p^{i}}{\p s_{i}}\right)\tilde{\nabla}_{a}s_{i}\;.
\ee
In near-perfect fluid analyses such as the present one, $\vep^{i}_{a}$ is often taken to be negligible. Thus in subsequent  discussions all terms containing this quantity are dropped.

\subsection{Curvature Fluid Variables}
The information about our deviation from standard GR is carried by the following  dimensionless gradient quantities 
\be
{\car}_{a}=a\tl \nb_{a}R\;,~~~~~~ \Re_{a}=a\tl \nb_{a}\dot{R}\;.
\ee
These variables describe the inhomogeneities in the Ricci scalar. Finally,  the velocity of the curvature fluid is defined, following \cite{BED}, by
\be 
V^{R}_{a}=-\frac{\tl \nb_{a}R}{\dot{R}}\;,
\ee
provided that $\dot{R}\neq 0$. Cases with constant scalar configurations or pathologic $f(R)$ models with points of inflection in $R(t)$ are excluded from this analysis.
%%%%%%%%%%%%%%%%%%%%%%%%%%%%%%%%%%%%%%%%%%%%%%%%%%%%%%%%%%
\section{Equations in the Energy Frame}
%%%%%%%%%%%%%%%%%%%%%%%%%%%%%%%%%%%%%%%%%%%%%%%%%%%%%%%%%%
We now derive the time evolution of the perturbations to linear order in the energy frame of the matter,  i.e.,  in the frame where $u^{a}=u^{a}_{m}$.
%%%%%%%%%%%%%%%%%%%%%%%%%%%%%%%%%%%%%%%%%%%%%%%%%%%%%%%%%%
\subsection{Total Fluid Equations}
%%%%%%%%%%%%%%%%%%%%%%%%%%%%%%%%%%%%%%%%%%%%%%%%%%%%%%%%%%
These equations characterize the temporal fluctuations of inhomogeneities in a generic perfect cosmological fluid with an equation of state  evolving as $\dot{w}=(1+w)(w-c^{2}_{s})\Theta$. They are the following: 

\be
\fl \dot{D}_{a}+(1+w)Z_{a}-w\Theta D_{a}=0\;,
\ee
\begin{eqnarray}
\fl\label{zone} \dot{Z}_{a}&-\left(\dot{R}\frac{f''}{f'}-\frac{2}{3}\Theta\right)Z_{a}-\left[ \frac{(1+3w)c^{2}_{s}-2(1+w)}{2(1+w)}\frac{\mu_{m}}{f'} +\frac{2c^{2}_{s}\Theta^{2}+3c^{2}_{s}(\mu_{R}+3p_{R})}{6(1+w)}\right.\nn
\fl  & \left.+\frac{c^{2}_{s}}{1+w}\frac{2K}{a^{2}}\right]D^{m}_{a}- \left[ \frac{2f'\Theta^{2}+3(1+3w)\mu_{m}+3f'(\mu_{R}+3p_{R})}{6f'(1+w)}+\frac{1}{1+w}\frac{2K}{a^{2}}\right] w\varepsilon_{a}\nn
\fl &-\Theta\frac{f''}{f'}\Re_{a}-\left[ \frac{1}{2}-\frac{1}{2}\frac{ff''}{f'^{2}}+\frac{f''\mu_{m}}{f'^{2}}-\dot{R}\Theta\left(\frac{f''}{f'}\right)^{2}+\dot{R}\Theta \frac{f'''}{f'}+\frac{2K}{a^{2}}\frac{f''}{f'}\right] {\car}_{a}\nn
\fl &+\frac{f''}{f'}\tl \nb^{2}{\car}_{a}+\frac{c^{2}_{s}\tl \nb^{2}D^{m}_{a}}{1+w}+\frac{w\tl \nb^{2}\vep_{a}}{1+w}=0\;,\\
\fl \dot{\car}_{a}&-\Re_{a}+\dot{R}\left[ \frac{c^{2}_{s}}{1+w}D_{a}+\frac{w}{1+w}\varepsilon_{a}\right]=0\;,\\
 \fl\dot{\Re}_{a}&+\left(2\dot{R}\frac{f'''}{f''}+\Theta\right)\Re_{a}-\left[\frac{(1-3c^{2}_{s})\mu_{m}}{3f''}-\frac{c^{2}_{s}}{1+w}\ddot{R}\right]D_{a}+\left(\frac{w\mu_{m}}{f''}+\frac{w}{1+w}\ddot{R}\right)\varepsilon_{a}\nn
\fl &+\dot{R}Z_{a}+\left(\frac{2K}{a^{2}}+\ddot{R}\frac{f'''}{f''}+\dot{R}^{2}\frac{f^{(iv)}}{f''}+\dot{R}\Theta \frac{f'''}{f''}
+\frac{1}{3}\frac{f'}{f''}-\frac{R}{3}\right)\car_{a}-\tilde{\nabla}^{2}\car_{a}=0\;.
\end{eqnarray}

The  linearized 3-curvature scalar of the projected metric $h_{ab}$ orthogonal to the 4-velocity vector $u^{a}$  is \cite{DBE} 
\be
\tl{R}=2\left(-\frac{1}{3}\Theta^{2}+\mu\right)
\ee     
reduces to the Ricci scalar in the hypersurfaces orthogonal to  $u^{a}$ when $\omega=0$.
The covariant, GI gradient $C_{a}$ gives, to linear order

\begin{eqnarray}
\fl \frac{C_{a}}{a^{2}} &+\left(\frac{4}{3}\Theta+2\frac{\dot{R}f''}{f'} \right)Z_{a}-2\frac{\mu_{m}}{f'}D^{m}_{a}+
2\Theta\frac{f''}{f'}\Re_{a}-2\frac{f''}{f'}\tl\nb^{2}{\car}_{a}\nn
\fl &+\left[2\Theta\dot{R}\frac{f'''}{f'}-\frac{f''}{f'}\left(\frac{f}{f'}-2\frac{\mu_{m}}{f'}+2\dot{R}\Theta\frac{f''}{f'}+\frac{4K}{a^{2}}\right)   \right] {\car}_{a}
=0\;.
\end{eqnarray}

This variable quantifies the spatial variation in the 3-curvature and is a geometrically natural quantity useful in the long wavelength analysis of our perturbation equations. The time evolution of this quantity is given by
\begin{eqnarray}
\fl  \dot{C}_{a}&=2K\left\{\frac{3f'C_{a}}{a^{2}(2\Theta f'+3\dot{R}f'')}+D^{m}_{a}\left[\frac{8w\Theta}{3(1+w)}-\frac{2f'\Theta^{2}-6f'\mu_{R}}{2\Theta f'+3\dot{R}f''}\right]-\frac{6f'' \tl\nb^{2}{\car}_{a}}{2\Theta f'+3\dot{R}f''}\right.\nn
\fl &\left.-\left[\frac{a^{2}(\Theta f''-3\dot{R}f''')}{3f'}\frac{12\dot{R}\Theta f'f'''-2f''\left(3f-2\Theta^{2}f'+6\Theta^{2}\mu_{R}+6\dot{R}\Theta f''\right)}{2\Theta f'+3\dot{R}f''}\right]{\car}_{a}\right.\nn
\fl &\left. +\left[\frac{6\Theta f''}{2\Theta f'+3\dot{R}f''}+\frac{f''}{f'}\right]\Re_{a}\right\}+K^{2}\left[\frac{36f''{\car}_{a}}{a^{2}(2\Theta f'+3\dot{R}f'')}-\frac{36f'D^{m}_{a}}{a^{2}(2\Theta f'+3\dot{R}f'')}\right]\nn
\fl & +\frac{2}{3}\tl\nb^{2}\left\{\frac{2wa^{2}\Theta}{(1+w)}D^{m}_{a}+\frac{a^2}{f'}\left[ 3f''\Re_{a}-(\Theta f''-3\dot{R}f''){\car}_{a}\right]\right\}\;.
\fl \label{cadot}
\end{eqnarray}
%%%%%%%%%%%%%%%%%%%%%%%%%%%%%%%%%%%%%%%%%%%%%%%%%%%%%%%%%%
\subsection{Component Equations}
%%%%%%%%%%%%%%%%%%%%%%%%%%%%%%%%%%%%%%%%%%%%%%%%%%%%%%%%%%
These are the equations that describe the evolution dynamics of the individual fluid component fluctuations.  For the component  matter and velocity fluctuations, these are given by
\begin{eqnarray}
\fl \dot{D}^{i}_{a}&-(w_{i}-c^{2}_{si})\Theta D^{i}_{a}+(1+w_{i})Z_{a}=\frac{1}{\mu_{i}h}h_{i}\Theta(c^{2}_{s}\mu D_{a}+p \varepsilon_{a})-a(1+w_{i})\tilde{\nabla}_{a}\tilde{\nabla}^{b}V^{i}_{b}\;,\\
\fl \dot{V}^{i}_{a}&-\left(c^{2}_{si}-\frac{1}{3}\right)\Theta V^{i}_{a}=\frac{1}{ahh_{i}}\left(h_{i}c^{2}_{s}\mu D_{a}+h_{i}p\varepsilon_{a}-hc^{2}_{si}\mu^{i} D^{i}_{a}\right)\;.
\end{eqnarray}
We note that the equations involving the gradients of the inhomogeneities in the expansion and curvature variables ($ Z_{a}, {\car}_{a},\Re_{a},C_{a}$) remain the same as in the total fluid equations (\ref{zone})-(\ref{cadot}). This is to be expected since  these quantities are global intrinsic properties of the spacetime itself rather than of the individual components of matter in the fluid. 
\subsection{Relative Equations}
Let us now define the variables that relate features of pairs of the different components of the fluid, and derive their governing evolution equations.
These relative variables depend only on the choice of the individual velocities, not on the choice of the overall frame.
\be
S^{ij}_{a} \equiv \frac{\mu_{i}D^{i}_{a}}{h_{i}}-\frac{\mu_{j}D^{j}_{a}}{h_{j}}\;,~~~~~~~~~~
V^{ij}_{a} \equiv V^{i}_{a}-V^{j}_{a}\;.
\ee
These are the quantities that allow us to distinguish between adiabatic and isothermal perturbations \cite{KS1984,DBE}.

The  derivation of the evolution equations for the above quantities is straightforward and yields
\begin{eqnarray}
\fl &\dot{V}^{ij}_{a}-(c^{2}_{sj}-\sfrac{1}{3})\Theta V^{ij}_{a}-(c^{2}_{si}-c^{2}_{sj})\Theta V^{i}_{a}=-\frac{1}{ah_{i}}(c^{2}_{si}-c^{2}_{sj})\mu_{i}D^{m}_{i}-\frac{1}{a}c^{2}_{sj}S^{ij}_{a}\;,\\
\fl &\dot{S}^{ij}_{a}+a\tilde{\nabla}_{a}\tilde{\nabla}^{b}V^{ij}_{b}=0\;.
\end{eqnarray}
%%%%%%%%%%%%%%%%%%%%%%%%%%%%%%%%%%%%%%%%%%%%%%%%%%%%%%%%%%
\section{Scalar Equations}
%%%%%%%%%%%%%%%%%%%%%%%%%%%%%%%%%%%%%%%%%%%%%%%%%%%%%%%%%%
The quantities we have considered so far contain both a scalar and a vector (solenoidal) part.  Structure formation on cosmological scales is believed to follow spherical clustering and therefore we present here the spherically symmetric, scalar density perturbations obtained by taking the divergence of the gradient quantities. In so doing, we first apply  a \textit{local decomposition} 
\be
a\tl\nb_{b}X_{a}=X_{ab}=\sfrac{1}{3}h_{ab}X+\Sigma^{X}_{ab}+X_{[ab]}\;, 
\ee
 where $\Sigma^{X}_{ab}=X_{(ab)}-\sfrac{1}{3}h_{ab}X$ describes shear whereas $X_{[ab]}$ describes the vorticity. 
 Vorticity and shear
describe the rotation and distortion of the density gradient 
  field, respectively. 
The above decomposition extracts the scalar part of the perturbation vectorial gradients.  Accordingly, when extracting the scalar contribution 
the vorticity term vanishes \cite{EBH}. 

\subsection{Scalar Variables}
On the basis of the above decomposition scheme, our scalar variables are:
\begin{eqnarray}
\fl &\del_{m}=a\tilde{\nabla}^{a}D^{m}_{a}\;,~~~~~~~~~~Z=a\tilde{\nabla}^{a}Z_{a}\;,~~~~~~~~~C=a\tilde{\nabla}^{a}C_{a}\;,~~~~~~\car=a\tilde{\nabla}^{a}\car_{a}\;, \nn
\fl &\Re=a\tilde{\nabla}^{a}\Re_{a}\;,~~~~~~~~~~~~~\varepsilon=a\tilde{\nabla}^{a}\varepsilon_{a}\;,~~~~~~~~~~~\del^{i}_{m}=a\tilde{\nabla}^{a}D^{i}_{a}\;,~~~~~~V_{i}=a\tilde{\nabla}^{a}V^{i}_{a}\;,\nn
\fl &S_{ij}=a\tilde{\nabla}^{a}S^{ij}_{a}\;,~~~~~~~~~~V_{ij}=a\tilde{\nabla}^{a}V^{ij}_{a}\;.
\end{eqnarray}
The scalar variables  describing the total fluid will thus evolve according to
\begin{eqnarray}
\fl \dot{\del}_{m}&+(1+w)Z-w\Theta \del_{m}=0\;,\\
\fl  \dot{Z} &-\left(\dot{R}\frac{f''}{f'}-\frac{2}{3}\Theta\right)Z-\left[ \frac{1}{2}-\frac{1}{2}\frac{ff''}{f'^{2}}+\frac{f''\mu_{m}}{f'^{2}}-\dot{R}\Theta\left(\frac{f''}{f'}\right)^{2}+\dot{R}\Theta \frac{f'''}{f'}\right] {\car}\nn
\fl & +\frac{f''}{f'}\tl \nb^{2}{\car}-\left[ \frac{(1+3w)c^{2}_{s}-2(1+w)}{2(1+w)}\frac{\mu_{m}}{f'} +\frac{2c^{2}_{s}\Theta^{2}+3c^{2}_{s}(\mu_{R}+3p_{R})}{6(1+w)}\right]\del_{m}\nn
\fl &- \left[ \frac{2f'\Theta^{2}+3(1+3w)\mu_{m}+3f'(\mu_{R}+3p_{R})}{6f'(1+w)}\right] w\varepsilon
-\Theta\frac{f''}{f'}\Re+\frac{c^{2}_{s}}{1+w}\tl \nb^{2}\del_{m}\nn
\fl &+\frac{w}{1+w}\tl \nb^{2}\vep=0\;,\\
\fl \dot{\car} &- \Re+\dot{R}\left( \frac{c^{2}_{s}}{1+w}\del_{m}+\frac{w}{1+w}\varepsilon\right)=0\;,
\end{eqnarray}
\begin{eqnarray}
\fl \dot{\Re}&+\left(2\dot{R}\frac{f'''}{f''}+\Theta\right)\Re+\left(\ddot{R}\frac{f'''}{f''}+\dot{R}^{2}\frac{f^{(iv)}}{f''}+\dot{R}\Theta \frac{f'''}{f''}
+\frac{1}{3}\frac{f'}{f''}-\frac{R}{3}\right)\car-\tilde{\nabla}^{2}\car\nn
\fl & +\dot{R}Z-\left[\frac{(1-3c^{2}_{s})\mu_{m}}{3f''}-\frac{c^{2}_{s}}{1+w}\ddot{R}\right]\del_{m}+\left[\frac{w\mu_{m}}{f''}+\frac{w}{1+w}\ddot{R}\right]\varepsilon=0\;,\\
\fl \frac{C}{a^{2}}&+\left(\frac{4}{3}\Theta+2\frac{\dot{R}f''}{f'} \right)Z+\left[2\Theta\dot{R}\frac{f'''}{f'}-\frac{f''}{f'}\left(\frac{f}{f'}-2\frac{\mu_{m}}{f'}+2\dot{R}\Theta\frac{f''}{f'}\right)\right] {\car}\nn
\fl &-2\frac{\mu_{m}}{f'}\del_{m}+2\Theta\frac{f''}{f'}\Re-2\frac{f''}{f'}\tl\nb^{2}{\car} =0\;.
\label{first1}
\end{eqnarray}
The evolution of the constraint equation is given by
\begin{eqnarray}
\fl \dot{C}&=K\left\{\frac{6f'C}{a^{2}(2\Theta f'+3\dot{R}f'')}+\del_{m}\left[\frac{16w\Theta}{3(1+w)}-\frac{4f'\Theta^{2}-12f'\mu_{R}}{2\Theta f'+3\dot{R}f''}\right]
 -\frac{12f'' \tl\nb^{2}{\car}}{2\Theta f'+3\dot{R}f''}\right.\nn
\fl &\left.~-\left[\frac{2a^{2}(\Theta f''-3\dot{R}f''')}{3f'}\frac{12\dot{R}\Theta f'f'''-2f''\left(3f-2\Theta^{2}f'+6\Theta^{2}\mu_{R}+6\dot{R}\Theta f''\right)}{2\Theta f'+3\dot{R}f''}\right]{\car}\right.\nn
\fl &\left.+\left(\frac{12\Theta f''}{2\Theta f'+3\dot{R}f''}+\frac{2f''}{f'}\right)\Re\right\}+ K^{2}\left[\frac{36f''{\car}}{a^{2}(2\Theta f'+3\dot{R}f'')}-\frac{36f'\del{m}}{a^{2}(2\Theta f'+3\dot{R}f'')}\right]\nn
\fl & +\tl\nb^{2}\left[\frac{4wa^{2}\Theta}{3(1+w)}\del_{m}+\frac{2a^{2}f''}{f'}\Re-\frac{2a^{2}(\Theta f''-3\dot{R}f'')}{3f'}{\car}\right]\;.
\label{cadots}
\end{eqnarray}
For the scalar variables describing  component inhomogeneities and interactions in the fluid, the evolution equations are given by
\begin{eqnarray}
\fl &\dot{\del}^{i}_{m}-\Theta(w_{i}-c^{2}_{si})\del^{i}_{m}+(1+w_{i})Z=\frac{1+w_{i}}{1+w}\left(c^{2}_{s}\Theta\del_{m}+w \Theta\varepsilon\right) -a(1+w_{i})\tilde{\nabla}^{2}V_{i}\;,\\
\fl &\dot{V}_{i}-\left(c^{2}_{si}-\frac{1}{3}\right)\Theta V_{i}=\frac{1}{ahh_{i}}\left(h_{i}c^{2}_{s}\mu \del_{m}+h_{i}p\varepsilon-hc^{2}_{si}\mu^{i}\del^{i}_{m} \right)\;,\\
\fl &\dot{V}_{ij}-\left(c^{2}_{sj}-\frac{1}{3}\right)\Theta V_{ij}-(c^{2}_{si}-c^{2}_{sj})\Theta V_{i}=-\frac{1}{ah_{i}}(c^{2}_{si}-c^{2}_{sj})\mu_{i}\del^{i}_{m}-\frac{1}{a}c^{2}_{sj}S_{ij}\;,\\
\fl &\dot{S}_{ij}+a\tilde{\nabla}^{2}V_{ij}=0\;.
\label{last1}
\end{eqnarray}
%%%%%%%%%%%%%%%%%%%%%%%%%%%%%%%%%%%%%%%%%%%%%%%%%%%%
\subsection{Second-order Equations}
%%%%%%%%%%%%%%%%%%%%%%%%%%%%%%%%%%%%%%%%%%%%%%%%%%%%
The above first order equations (\ref{first1})-(\ref{last1}) can be reduced to  a set of linearly independent second order equations. This has the advantage of simplifying the equations  and making comparisons to GR more transparent \cite{CDT}:
\begin{eqnarray}
\fl \ddot{\del}_{m} &+\left[ (c^{2}_{s}+\frac{2}{3}-2w)\Theta -\dot{R}\frac{f''}{f'}\right] \dot{\del}_{m}-c^{2}_{s}\tl\nb^{2}\del_{m}+\left[\left(\frac{3}{2}w^{2}+5c^{2}_{s}-4w-1\right)\frac{\mu_{m}}{f'}
 \right.\nn
\fl  &\left. +\frac{1}{2}(3w-5c^{2}_{s})\frac{f}{f'} +(c^{2}_{s}-w)\left( 2R-4\dot{R}\Theta\frac{f''}{f'}-\frac{12K}{a^{2}}\right)\right]\del_{m} -(1+w)\frac{f''}{f'}\tl\nb^{2}{\car}\nn
\fl &=\frac{1+w}{2}\left[ -1+(f-2\mu_{m}+2\dot{R}\Theta f'')\frac{f''}{f'^{2}}-2\dot{R}\Theta\left(\frac{f''}{f'}\right)^{2}
-2\dot{R}\Theta\frac{f'''}{f'}  \right] \car\nn
\fl &-(1+w)\Theta \frac{f''}{f'}\dot{\car}-\left(2\frac{\mu_{m}}{f'}+ \frac{R}{2}-\frac{f}{f'}-\dot{R}\Theta\frac{f''}{f'}-\frac{3K}{a^{2}}\right)w\vep+w\tl\nb^{2}\vep
\label{first2}\;,
\end{eqnarray}
\begin{eqnarray}
\fl \ddot{\car} &+\left(2\dot{R}\frac{f'''}{f''}+\Theta\right)\dot{\car}+\left(\ddot{R}\frac{f'''}{f''}+\dot{R}^{2}\frac{f^{(iv)}}{f''}+\dot{R}\Theta \frac{f'''}{f''}
+\frac{1}{3}\frac{f'}{f''}-\frac{R}{3}\right)\car\nn
\fl & +\frac{c^{2}_{s}-1}{1+w}\dot{R}\dot{\del}_{m}+\left\{\frac{(3c^{2}_{s}-1)\mu_{m}}{3f''}+\frac{w+c^{2}_{s}}{1+w}\dot{R}\Theta+\frac{2c^{2}_{s}}{1+w}\left(\ddot{R}+\dot{R}^{2}\frac{f'''}{f''}\right)\right.\nn
\fl &\left. +\frac{\dot{R}}{1+w}\left[\dot{c^{2}_{s}}+c^{2}_{s}(c^{2}_{s}-w)\Theta\right]\right\} \del_{m}
-\tl\nb^{2}\car+\frac{w}{1+w}\dot{R}\dot{\vep}\nn
\fl &+\left[\frac{w\mu_{m}}{f''}+\frac{2w-c^{2}_{s}}{1+w}\dot{R}\Theta+\frac{2w}{1+w}\left(\ddot{R}+\dot{R}^{2}\frac{f'''}{f''}\right)\right]\vep=0\;, \\
\fl \ddot{\del}_{i} &+ \left(\frac{2}{3}-w_{i}\right)\Theta\dot{\del}_{i} -\frac{1+w_{i}}{1+w}\left[\dot{R}\frac{f''}{f'}+ (c^{2}_{s}-c^{2}_{si})\Theta\right]\dot{\del}_{m}-(1+w_{i})\frac{f''}{f'}\tl\nb^{2}\car\nn
\fl &-\frac{1+w_{i}}{1+w}\left[(1+w)\frac{\mu_{m}}{f'}
-\left(2\frac{\mu_{m}}{f'}-\frac{f}{f'}-2\Theta\dot{R}\frac{f''}{f'}\right)c^{2}_{s}+\dot{c^{2}_{s}}\Theta\right.\nn
\fl &\left. +(c^{2}_{s}-c^{2}_{si})(c^{2}_{s}-w)\Theta^{2}
-(c^{2}_{s}+w)\dot{R}\Theta \frac{f''}{f'} \right]\del_{m}-\frac{1+w_{i}}{1+w}w\Theta \dot{\vep} -c^{2}_{si}\tl\nb^{2}\del_{i}\nn
\fl &+(1+w_{i})\Theta \frac{f''}{f'}\dot{\car}
+(1+w_{i}) \left[ \frac{1}{2}-\frac{1}{2}\frac{ff''}{f'^{2}}+\frac{f''\mu_{m}}{f'^{2}}
-\dot{R}\Theta\left(\frac{f''}{f'}\right)^{2}+\dot{R}\Theta\frac{f'''}{f'} \right] \car\nn
\fl &-\frac{1+w_{i}}{1+w}\left[(w-c^{2}_{s}-c^{2}_{si}w)\Theta^{2} -w\left(2\frac{\mu_{m}}{f'}-\frac{f}{f'}-\dot{R}\Theta\frac{f''}{f'}\right)\right]\vep=0\;.
\label{last2}
\end{eqnarray}

The second order equations (\ref{first2})-(\ref{last2}) governing the propagation of the entropy perturbations for a general $\vep$ (or $S_{ij}$) are in general very complicated and consequently we will present them for specific (radiation-dust) applications in section \ref{sec:dustrad}.

%%%%%%%%%%%%%%%%%%%%%%%%%%%%%%%%%%%%%%%%%%%%%%%%%%%%
 \section{Harmonic Analysis}
 %%%%%%%%%%%%%%%%%%%%%%%%%%%%%%%%%%%%%%%%%%%%%%%%%%%%
The above evolution equations can be thought of as a coupled system of harmonic oscillators of the form
\be
\ddot{X}+A\dot{X}+BX=C(Y,\dot{Y})\;,
\ee
where the second term  from the left represents the friction (damping) term, the third one, the restoring force term while $C$ represents the source forcing term. A key assumption in the analysis of the equation here is that we can apply the separation of variables technique
\be
X(x,t)=X(\vec{x})X(t)\;,~~~~Y(x,t)=Y(\vec{x})Y(t)\;,
\ee
and write
\be
X=\sum_{k}{X^{k}(t)}Q_{k}(\vec{x})\;,~~~~~~
Y=\sum_{k}{Y^{k}(t)}Q_{k}(\vec{x})\;,
\ee
where $Q_{k}(x)$ are the eigenfunctions of the  covariantly defined Laplace-Beltrami operator on an almost FLRW space-time:  
\be 
\tlnb Q=-\frac{k^{2}}{a^{2}}Q\;.
\ee
Here $k=\frac{2\pi a}{\lambda}$ is the order of the harmonic and $\dot{Q}_{k}(\vec{x})=0$ (Q is covariantly constant). In this way the evolution equations and the constraint equation can be converted into ordinary differential equations for each mode. After harmonic decomposition the first order total and component fluid equations (\ref{first1})-(\ref{last1}) can be rewritten in the following form:
\begin{eqnarray}
\fl &\dot{\del}^{k}_{m}+(1+w)Z^{k}-w\Theta \del^{k}_{m}=0\;,\\
\fl & \dot{Z}^{k}-\left(\dot{R}\frac{f''}{f'}-\frac{2}{3}\Theta\right)Z^{k}\nn 
\fl&-\left[ \frac{(1+3w)c^{2}_{s}-2(1+w)}{2(1+w)}\frac{\mu_{m}}{f'} +\frac{2c^{2}_{s}\Theta^{2}+3c^{2}_{s}(\mu_{R}+3p_{R})}{6(1+w)}+\frac{c^{2}_{s}}{1+w}\frac{k^{2}}{a^{2}}\right]\del^{k}_{m}\nn
\fl &- \left[ \frac{2f'\Theta^{2}+3(1+3w)\mu_{m}+3f'(\mu_{R}+3p_{R})}{6f'(1+w)}+\frac{1}{1+w}\frac{k^{2}}{a^{2}}\right] w\vep^{k}
-\Theta\frac{f''}{f'}\Re^{k}\nn
\fl &-\left[ \frac{1}{2}+\frac{k^{2}}{a^{2}}\frac{f''}{f'}-\frac{1}{2}\frac{ff''}{f'^{2}}+\frac{f''\mu_{m}}{f'^{2}}-\dot{R}\Theta\left(\frac{f''}{f'}\right)^{2}+\dot{R}\Theta \frac{f'''}{f'}\right] {\car}^{k}=0,\\
\fl &\dot{\car}^{k}- \Re^{k}+\dot{R}\left(\frac{c^{2}_{s}}{1+w}\del^{k}_{m}+\frac{w}{1+w}\vep^{k}\right)=0\;,\\
\fl &\dot{\Re}^{k}+\left(2\dot{R}\frac{f'''}{f''}+\Theta\right)\Re^{k}+\left[\frac{(1-3c^{2}_{s})\mu_{m}}{3f''}-\frac{c^{2}_{s}}{1+w}\ddot{R}\right]\del^{k}_{m}+\left[\frac{w\mu_{m}}{f''}+\frac{w}{1+w}\ddot{R}\right]\vep^{k}\nn
\fl &+\dot{R}Z^{k}+\left(\frac{k^{2}}{a^{2}}+\ddot{R}\frac{f'''}{f''}+\dot{R}^{2}\frac{f^{(iv)}}{f''}+\dot{R}\Theta \frac{f'''}{f''}
+\frac{1}{3}\frac{f'}{f''}-\frac{R}{3}\right)\car^{k}=0\;,\\
\fl & \frac{C^{k}}{a^{2}}+\left(\frac{4}{3}\Theta+2\frac{\dot{R}f''}{f'} \right)Z^{k}-2\frac{\mu_{m}}{f'}\del^{k}_{m}\nn
\fl & +\left[2\Theta\dot{R}\frac{f'''}{f'}-\frac{f''}{f'}\left(\frac{f}{f'}-2\frac{\mu_{m}}{f'}+2\dot{R}\Theta\frac{f''}{f'}-2\frac{k^{2}}{a^{2}}\right)\right] {\car}
+2\Theta\frac{f''}{f'}\Re^{k}=0\;,
\end{eqnarray}
\begin{eqnarray}
\label{cadotk}
\fl \dot{C}^{k}&=K^{}\left[\frac{36K(f''{\car^{k}}-f'\del^{k}_{m})+6f'C^{k}}{a^{2}(2\Theta f'+3\dot{R}f'')}\right] +K\left(\frac{12\Theta f''}{2\Theta f'+3\dot{R}f''}+\frac{2f''}{f'}\right)\Re\nn
\fl &+K\left\{\del^{k}_{m}\left[\frac{16w\Theta}{3(1+w)}-\frac{4f'\Theta^{2}-12f'\mu_{R}}{2\Theta f'+3\dot{R}f''}\right]+\frac{12f''}{2\Theta f'+3\dot{R}f''}\frac{k^{2}}{a^{2}}{\car^{k}}\right.\nn
\fl &\left.-\frac{2 {\car^{k}}a^{2}(\Theta f''-3\dot{R}f''')}{3f'}\frac{12\dot{R}\Theta f'f'''-2f''\left(3f-2f'(\Theta^{2}-3\mu_{R})+6\dot{R}\Theta f''\right)}{2\Theta f'+3\dot{R}f''}\right\}\nn
\fl &-\frac{k^{2}}{a^{2}}\left[\frac{4wa^{2}\Theta}{3(1+w)}\del_{m}+\frac{2a^{2}f''}{f'}\Re^{k}-\frac{2a^{2}(\Theta f''-3\dot{R}f'')}{3f'}{\car^{k}}\right]\;,
\end{eqnarray}
\begin{eqnarray}
\fl &\dot{\del}^{k}_{i}-(w_{i}-c^{2}_{si})\Theta\del^{k}_{i}+(1+w_{i})Z^{k}=\frac{1+w_{i}}{1+w} \left(c^{2}_{s}\del^{k}_{m}+w \vep^{k}\right)\Theta +(1+w_{i})\frac{k^{2}}{a}V_{i}\;,\\
\fl &\dot{V}^{k}_{i}-\left(c^{2}_{si}-\frac{1}{3}\right)\Theta V^{k}_{i}=\frac{1}{ahh_{i}}\left(h_{i}c^{2}_{s}\mu \del_{m}+h_{i}p\vep^{k}-hc^{2}_{si}\mu^{i}\del^{k}_{i} \right) ,\\
\fl &\dot{V}^{k}_{ij}-\left(c^{2}_{sj}-\frac{1}{3}\right)\Theta V^{k}_{ij}-(c^{2}_{si}-c^{2}_{sj})\Theta V^{k}_{i}=-\frac{1}{ah_{i}}(c^{2}_{si}-c^{2}_{sj})\mu_{i}\del^{k}_{i}-\frac{1}{a}c^{2}_{sj}S^{k}_{ij}\;,\\
\fl &\dot{S}^{k}_{ij}-\frac{k^{2}}{a}V^{k}_{ij}=0\;.
\end{eqnarray}
\normalsize
The  form and use of Eqn. (\ref{cadotk}) will be more transparent when we discuss the long wavelength limits of our perturbations for radiation and dust backgrounds. The harmonically decomposed set of second-order equations (\ref{first2})-(\ref{last2}) will become
\begin{eqnarray}
\fl \ddot{\del}^{k}_{m}&+\left[ (c^{2}_{s}+\frac{2}{3}-2w)\Theta -\dot{R}\frac{f''}{f'}\right] \dot{\del}^{k}_{m}+\left[\left(\frac{3}{2}w^{2}+5c^{2}_{s}-4w-1\right)\frac{\mu_{m}}{f'}
 \right.\nn
 \fl &\left. +\frac{1}{2}(3w-5c^{2}_{s})\frac{f}{f'} +(c^{2}_{s}-w)\left( 2R-4\dot{R}\Theta\frac{f''}{f'}-\frac{12K}{a^{2}}\right)+c^{2}_{s}\frac{k^{2}}{a^{2}}\right]\del^{k}_{m} \nn
\fl &=\frac{1+w}{2}\left[ -1-\frac{2k^{2}}{a^{2}}\frac{f''}{f'}+(f-2\mu_{m}+2\dot{R}\Theta f'')\frac{f''}{f'^{2}}-2\dot{R}\Theta\left((\frac{f''}{f'})^{2}
+\frac{f'''}{f'}\right)  \right] \car^{k}\nn
\fl &-(1+w)\Theta \frac{f''}{f'}\dot{\car}^{k}-\left[2\frac{\mu_{m}}{f'}+ \frac{R}{2}-\frac{f}{f'}-\dot{R}\Theta\frac{f''}{f'}-\frac{3K}{a^{2}}+\frac{k^{2}}{a^{2}}\right] w\varepsilon^{k}\;, \label{first3}
\end{eqnarray}
\begin{eqnarray}
\fl \ddot{\car}^{k}&+\left(2\dot{R}\frac{f'''}{f''}+\Theta\right)\dot{\car}^{k}+\left(\frac{k^{2}}{a^{2}}+\ddot{R}\frac{f'''}{f''}+\dot{R}^{2}\frac{f^{(iv)}}{f''}+\dot{R}\Theta \frac{f'''}{f''}
+\frac{1}{3}\frac{f'}{f''}-\frac{R}{3}\right)\car^{k}\nn
\fl &+\frac{c^{2}_{s}-1}{1+w}\dot{R}\dot{\del}^{k}_{m} +\left[\frac{w\mu_{m}}{f''}+\frac{2w-c^{2}_{s}}{1+w}\dot{R}\Theta+\frac{w}{1+w}\left(2\ddot{R}+2\dot{R}^{2}\frac{f'''}{f''}\right)\right]\vep^{k}\nn
\fl &+\left\{\frac{(3c^{2}_{s}-1)\mu_{m}}{3f''}+\frac{w+c^{2}_{s}}{1+w}\dot{R}\Theta+\frac{c^{2}_{s}}{1+w}\left(2\ddot{R}+2\dot{R}^{2}\frac{f'''}{f''}\right)\right.\nn
\fl & \left. +\frac{\dot{R}}{1+w}\left[\dot{c^{2}_{s}}+c^{2}_{s}(c^{2}_{s}-w)\Theta\right]\right\}\del^{k}_{m} +\frac{w}{1+w}\dot{R}\dot{\vep}^{k}=0\;,\\
\fl \ddot{\del}^{k}_{i}&+ (\frac{2}{3}-w_{i})\Theta\dot{\del}^{k}_{i} -\frac{1+w_{i}}{1+w}\left[\dot{R}\frac{f''}{f'}+ \left(c^{2}_{s}-c^{2}_{si}\right)\Theta\right]\dot{\del}^{k}+(1+w_{i})\Theta \frac{f''}{f'}\dot{\car}^{k}\nn
\fl &-\frac{1+w_{i}}{1+w}\left[(1+w)\frac{\mu_{m}}{f'}
-\left(2\frac{\mu_{m}}{f'}-\frac{f}{f'}-2\Theta\dot{R}\frac{f''}{f'}\right)c^{2}_{s}+\dot{c^{2}_{s}}\Theta\right.\nn
\fl &\left.+(c^{2}_{s}-c^{2}_{si})(c^{2}_{s}-w)\Theta^{2}
-(c^{2}_{s}+w)\dot{R}\Theta \frac{f''}{f'} \right]\del^{k}
-\frac{1+w_{i}}{1+w}w\Theta \dot{\vep}^{k}\nn
\fl &-\frac{1+w_{i}}{1+w}\left[(w-c^{2}_{s}-c^{2}_{si}w)\Theta^{2} -w\left(2\frac{\mu_{m}}{f'}-\frac{f}{f'}-\dot{R}\Theta\frac{f''}{f'}\right)\right]\vep^{k}+c^{2}_{si}\frac{k^{2}}{a^{2}} \del^{k}_{i}\nn
\fl &+(1+w_{i}) \left[ \frac{1}{2}+\frac{k^{2}}{a^{2}}\frac{f''}{f'}-\frac{1}{2}\frac{ff''}{f'^{2}}+\frac{f''\mu_{m}}{f'^{2}}
 -\dot{R}\Theta \left(\frac{f''}{f'}\right)^{2}+\dot{R}\Theta\frac{f'''}{f'} \right] \car^{k}=0\;.\label{last3}
 \end{eqnarray}
 As can be seen, this second order set of equations is not closed. For a two-component fluid the entropy and velocity perturbations equations are given by
 \begin{eqnarray}
\fl  \label{sddot}\ddot{S}^{k}_{ij}&=\frac{k^{2}}{a}\dot{V}_{ij}-\frac{k^{2}}{3a}\Theta V_{ij}\;,\\
\fl \label{vddot}\ddot{V}^{k}_{ij}&=\left(c^{2}_{z}-\frac{1}{3}\right)\Theta\dot{V}_{ij}+\frac{c^{2}_{si}-c^{2}_{sj}}{a(1+w)}\left(\frac{1}{3}+w-c^{2}_{s}\right)\Theta\del_{m}
-\frac{c^{2}_{z}}{a}\dot{S}_{ij}+\frac{c^{2}_{z}\Theta-3\dot{c}^{2}_{z}}{3a} S_{ij}\nn
\fl & +\left\{\dot{c}^{2}_{z}\Theta-\left(c^{2}_{z}-\frac{1}{3}\right)\left[\frac{1}{3}\Theta^{2}+\frac{1}{2}(1+3w)\frac{\mu_{m}}{f'}+\frac{1}{2}(\mu_{R}+3p_{R})\right]\right\} V_{ij}-\frac{c^{2}_{si}-c^{2}_{sj}}{a(1+w)}\dot{\del}_{m}\;.\nn
\fl&
\end{eqnarray}

Since Eqns. (\ref{sddot}) and (\ref{vddot}) are not linearly independent equations, we can choose either one of them  to close our system of second order equations (\ref{first3}-\ref{last3}).
%%%%%%%%%%%%%%%%%%%%%%%%%%%%%%%%%%%%%%%%%%%%%%%%%%%%
\section{Perturbations in a Radiation\hs Dust Universe}\label{sec:dustrad}
%%%%%%%%%%%%%%%%%%%%%%%%%%%%%%%%%%%%%%%%%%%%%%%%%%%%
\subsection{Background Setup}
%%%%%%%%%%%%%%%%%%%%%%%%%%%%%%%%%%%%%%%%%%%%%%%%%%%%
Now that we have derived the equations for perturbations of a general multi-fluid system, we consider an application of the equations for  a cosmological medium containing a non-interacting radiation-dust mixture and described by a flat ($K=0$) FLRW spacetime. Since our component fluids satisfy the conservation equations separately, we write
\begin{eqnarray}
&\dot{\mu}_{d}+\Theta\mu_{d}=0\label{dust}\;,\\
&\dot{\mu}_{r}+\frac{4}{3}\Theta\mu_{r}=0\label{rad}\;,
\end{eqnarray}
where $d$ and $r$ subindices hold for dust and radiation respectively.

The general equation of state $w$ for such a radiation-dust mixture is given by
\be
w=\frac{p_{m}}{\mu_{m}}=\frac{p_{d}+p_{r}}{\mu_{d}+\mu_{r}}=\frac{1}{3}\frac{\mu_{r}}{\mu_{d}+\mu_{r}}
\ee
and the speed of sound in the mixture is
\be
c^{2}_{s}=\frac{\dot{p}_{m}}{\dot{\mu}_{m}}=\frac{4\mu_{r}}{3(3\mu_{d}+4\mu_{r})}\;.
\ee
Wherever necessary, we will use the shorthand
\begin{equation}
c^{2}_{z}\equiv\frac{1}{h}\left(h_{r}c^{2}_{sd}+h_{d}c^{2}_{sr}\right)\;.
\end{equation}

In general, since we do not have an explicit expression of the Hubble parameter $H$  and the curvature scalar $R$ as  functions of 
the scale factor $a$ in generic $f(R)$ gravity theories, an exact multi-fluid background solution is not available and numerical solutions need to be obtained. This important issue will be investigated in a future work. 
 
 In this paper, we will confine our discussion to $R^{n}$  models  \cite{CDCT, CDT} and look for solutions in the short wavelength and long wavelength approximations for perturbations deep in the radiation and dust dominated epochs. During these epochs, since one fluid is negligible with respect to the other,  we 
 can use the exact single fluid background transient solution for $R^n$ models given by 
 $a=a_{eq}(t/t_{eq})^{\frac{2n}{3(1+w)}}$ where $a_{eq}$ is the scale factor at the time of radiation-dust equality $t_{eq}$ and will henceforth be normalized to unity.

In $R^n$ models the expressions for the expansion, the Ricci scalar, the curvature fluid energy density, the curvature fluid pressure and the effective matter energy density are given by
\begin{eqnarray}\label{theta}
\fl &\Theta=\frac{2n}{(1+w)t}\;,\\
\fl &R=\frac{4n\left[4n-3(1+w)\right]}{3(1+w)^{2}t^{2}}\;,\\
\fl &\mu_{R}=\frac{2(n-1)\left[2n(3w+5)-3(1+w)\right]}{3(1+w)^{2}t^{2}}\;,\\
\fl &p_{R}=\frac{2(n-1)\left[n(6w^{2}+8w-2)-3w(1+w)\right]}{3(1+w)^{2}t^{2}}\;,\\ \label{mur}
\fl &\mu_{m}=\left(\frac{3}{4}\right)^{1-n}\left[\frac{4n^2-3n(1+w)}{(1+w)^{2}t^{2}}\right]^{n-1}\frac{4n^{3}-2n(n-1)\left[2n(3w+5)-3(1+w)\right]}{3(1+w)^{2}t^{2}}\;.
\end{eqnarray}
%%%%%%%%%%%%%%%%%%%%%%%%%%%%%%%%%%%%%%%%%%%%%%%%%%%%
\subsection{Total Fluid Equations}
%%%%%%%%%%%%%%%%%%%%%%%%%%%%%%%%%%%%%%%%%%%%%%%%%%%%
Upon expanding Eqn. (\ref{pvep}) for a  mixture of dust and radiation, we obtain
\be
\fl p_{m}\varepsilon=-\frac{4\mu_{d}\mu_{r}}{3(3\mu_{d}+4\mu_{r})}S_{dr}\;,
\ee
and hence
\be
\fl \varepsilon=-\frac{4\mu_{d}}{3\mu_{d}+4\mu_{r}}S_{dr}\;.
\ee
We can thus readily derive the evolution equation for $\vep$ as follows
\be
\fl \dot{\varepsilon}=-\frac{16\mu_{d}\mu_{r}\Theta}{3(3\mu_{d}+4\mu_{r})^{2}}S_{dr}-\frac{4\mu_{d}}{3\mu_{d}+4\mu_{r}}\dot{S}_{dr}=-4c^{2}_{z}c^{2}_{s}\Theta S_{dr}-4c^{2}_{z}\dot{S}_{dr}\;.
\ee

Using these relations and applying the general total fluid second order equations to the radiation-dust mixture yields

\begin{eqnarray}
\fl \ddot{\del}^{k}_{m}&+\left[ \left(c^{2}_{s}+\frac{2}{3}-2w\right)\Theta -\dot{R}\frac{f''}{f'}\right] \dot{\del}^{k}_{m}-4wc^{2}_{z}\left[2\frac{\mu_{m}}{f'}+ \frac{R}{2}-\frac{f}{f'}-\dot{R}\Theta\frac{f''}{f'}+\frac{k^{2}}{a^{2}}\right]S^{k}_{dr}\nn
\fl &+\left[\left(\frac{3}{2}w^{2}+5c^{2}_{s}-4w-1\right)\frac{\mu_{m}}{f'}
 +\frac{1}{2}(3w-5c^{2}_{s})\frac{f}{f'} +(c^{2}_{s}-w)\left( 2R-4\dot{R}\Theta\frac{f''}{f'}\right)\right.\nn
 \fl &\left.+c^{2}_{s}\frac{k^{2}}{a^{2}}\right]\del^{k}_{m} -\frac{1+w}{2}\left[ -1-\frac{2k^{2}}{a^{2}}\frac{f''}{f'}+(f-2\mu_{m}+2\dot{R}\Theta f'')\frac{f''}{f'^{2}}-2\dot{R}\Theta \left(\frac{f''}{f'}\right)^{2}\right.\nn
\fl &\left.-2\dot{R}\Theta\frac{f'''}{f'}  \right] \car^{k}+(1+w)\Theta \frac{f''}{f'}\dot{\car}^{k}=0\;,
\end{eqnarray}
\begin{eqnarray}
\fl \ddot{\car}^{k}&+\left(2\dot{R}\frac{f'''}{f''}+\Theta\right)\dot{\car}^{k}+\left[\frac{k^{2}}{a^{2}}+\ddot{R}\frac{f'''}{f''}+\dot{R}^{2}\frac{f^{(iv)}}{f''}+\dot{R}\Theta \frac{f'''}{f''}
+\frac{1}{3}\frac{f'}{f''}-\frac{R}{3}\right]\car^{k}\nn
\fl &+\frac{c^{2}_{s}-1}{1+w}\dot{R}\dot{\del}^{k}_{m}+\left\{\frac{(3c^{2}_{s}-1)\mu_{m}}{3f''}+\frac{w+c^{2}_{s}}{1+w}\dot{R}\Theta+\frac{c^{2}_{s}}{1+w}\left(2\ddot{R}+2\dot{R}^{2}\frac{f'''}{f''}\right)\right.\nn
\fl &\left.+\frac{\dot{R}}{1+w}\left[\dot{c^{2}_{s}}+c^{2}_{s}(c^{2}_{s}-w)\Theta\right]\right\}\del^{k}_{m}-4wc^{2}_{z}\left[\frac{2}{1+w}\left(\ddot{R}+\dot{R}\Theta+\dot{R}^{2}\frac{f'''}{f''}\right)\right.\nn
\fl &\left.+\frac{\mu_{m}}{f''}\right]S^{k}_{dr}-\frac{4w}{1+w}c^{2}_{z}\dot{R}\dot{S}^{k}_{dr}=0\;, \\
\fl &\ddot{S}^{k}_{dr}+\left(c^2_{s}+\frac{1}{3}\right)\Theta \dot{S}^{k}_{dr}+\frac{k^{2}}{a^{2}}c^2_{z}S^{k}_{dr}-\frac{k^{2}}{a^{2}}\left(c^2_{z}+\frac{3}{4}c^{2}_{s}\right)\del^{k}_{m}=0\;,
\end{eqnarray} 
where $\del_{m}$ and $S_{dr}$ are given by $ \del_{m}=\frac{\mu_{d}\del_{d}+\mu_{r}\del_{r}}{\mu_{d}+\mu_{r}}\;,  \;S_{dr}=\del_{d}-\sfrac{3}{4}\del_{r}\;. $
%%%%%%%%%%%%%%%%%%%%%%%%%%%%%%%%%%%%%%%%%%%%%%%%%%%%
\subsection{Component Equations}
%%%%%%%%%%%%%%%%%%%%%%%%%%%%%%%%%%%%%%%%%%%%%%%%%%%%
The perturbations of the density gradients of the radiation component of the fluid are described by the  propagation equation
\begin{eqnarray}
\fl \ddot{\del}^{k}_{r}&+\left\{\left[\frac{1}{3}-\frac{4}{3(1+w)}\left(\frac{3w\mu_{d}}{3\mu_{d}+4\mu_{r}}+\frac{(c^{2}_{s}-\frac{1}{3})\mu_{r}}{\mu_{d}+\mu_{r}}\right)\right]\Theta-\frac{4\mu_{r}\dot{R}f''/f'}{3(1+w)(\mu_{d}+\mu_{r})}\right\}\dot{\del}^{k}_{r}\nn
\fl &+\frac{4\mu_{d}}{3(1+w)}
\left[\left(\frac{4w}{3\mu_{d}+4\mu_{r}}-\frac{c^{2}_{s}-\frac{1}{3}}{\mu_{d}+\mu_{r}}\right)\Theta-\frac{\dot{R}f''}{(\mu_{d}+\mu_{r})f'}\right]\dot{\del}^{k}_{d}\nn
\fl &+\frac{4}{3(1+w)}
\left[\frac{k^{2}}{3a^{2}}+\left(\frac{(w-c^{2}_{s})\mu_{r}}{\mu_{d}+\mu_{r}}-\frac{3w\mu_{d}}{3\mu_{d}+4\mu_{r}}+
\frac{\mu_{d}\mu_{r}}{3(\mu_{d}+\mu_{r})^{2}}\right)\dot{R}\Theta \frac{f''}{f'}\right.\nn
\fl &-\left.\left(\frac{4\mu_{d}\mu_{r}}{(3\mu_{d}+4\mu_{r})^{2}}\frac{3w\mu_{d}+(3w-1)\mu_{r}}{3(\mu_{d}+\mu_{r})}-\frac{3(c^{2}_{s}-\frac{2w}{3})\mu_{d}}{3\mu_{d}+4\mu_{r}}+
\frac{(c^{2}_{s}-\frac{1}{3})(c^{2}_{s}-w)\mu_{r}}{\mu_{d}+\mu_{r}}
\right.\right.\nn
\fl &\left.\left.-\frac{(c^{2}_{s}-\frac{1}{3})\mu_{d}\mu_{r}}{3(\mu_{d}+\mu_{r})^{2}}\right)\Theta^{2}
-\left(\frac{\left(1+w-2c^{2}_{s}\right)\mu_{r}}{\mu_{d}+\mu_{r}}-
\frac{6w\mu_{d}}{3\mu_{d}+4\mu_{r}}\right)\frac{\mu_{d}+\mu_{r}}{f'}\right.\nn
\fl &\left.-\left(\frac{3w\mu_{d}}{3\mu_{d}+4\mu_{r}}+\frac{c^{2}_{s}\mu_{r}}{\mu_{d}+\mu_{r}}\right)\frac{f}{f'}\right]\del^{k}_{r}+\frac{4}{3(1+w)}\left[\left(\frac{(w-c^{2}_{s})\mu_{d}}{\mu_{d}+\mu_{r}}+\frac{4w\mu_{d}}{3\mu_{d}+4\mu_{r}}\right.\right.\nn
\fl &\left.\left.-\frac{\mu_{d}\mu_{r}}{3(\mu_{d}+\mu_{r})^{2}}\right)\dot{R}\Theta \frac{f''}{f'}+\left(\frac{4\mu_{d}\mu_{r}}{3(3\mu_{d}+4\mu_{r})^{2}} \frac{4w\mu_{r}+(4w+1)\mu_{d}}{\mu_{d}+\mu_{r}}
-\frac{(c^{2}_{s}-\frac{1}{3})\mu_{d}\mu_{r}}{3(\mu_{d}+\mu_{r})^{2}}\right.\right.\nn
\fl &\left.\left.-\frac{4(c^{2}_{s}-\frac{2w}{3})\mu_{d}}{3\mu_{d}+4\mu_{r}}-\frac{(c^{2}_{s}-\frac{1}{3})(c^{2}_{s}-w)\mu_{d}}{\mu_{d}+\mu_{r}}\right)\Theta^{2}+\left(\frac{4w\mu_{d}}{3\mu_{d}+4\mu_{r}}-\frac{c^{2}_{s}\mu_{d}}{\mu_{d}+\mu_{r}}\right)\frac{f}{f'}\right.\nn
\fl &\left. -\left(\frac{\left(1+w-2c^{2}_{s}\right)\mu_{d}}{\mu_{d}+\mu_{r}}+
\frac{8w\mu_{d}}{3\mu_{d}+4\mu_{r}}\right)\frac{\mu_{d}+\mu_{r}}{f'}\right]\del^{k}_{d}+\frac{4}{3}\Theta \frac{f''}{f'}\dot{\car}^{k}\nn
\fl &+\frac{4}{3} \left[ \frac{1}{2}+\frac{k^{2}}{a^{2}}\frac{f''}{f'}-\frac{1}{2}\frac{ff''}{f'^{2}}+\frac{f''(\mu_{r}+\mu_{d})}{f'^{2}}
 -\dot{R}\Theta \left(\frac{f''}{f'}\right)^{2}+\dot{R}\Theta\frac{f'''}{f'} \right] \car^{k}=0\;.
 \fl
 \end{eqnarray}
 
 Similarly the propagation equation of the dust density gradient is given by\\
 \begin{eqnarray}
\fl \ddot{\del}^{k}_{d}&+\left[\left(\frac{2}{3}+\frac{\mu_{d}}{1+w}(\frac{4w}{3\mu_{d}+4\mu_{r}}-\frac{c^{2}_{s}}{\mu_{d}+\mu_{r}})\right)\Theta-\frac{\mu_{d}}{(1+w)(\mu_{d}+\mu_{r})}\dot{R}\frac{f''}{f'}\right]\dot{\del}^{k}_{d} \nn
\fl &-\frac{1}{1+w}\left[\left(\frac{3w\mu_{d}}{3\mu_{d}+4\mu_{r}}+\frac{c^{2}_{s}\mu_{r}}{\mu_{d}+\mu_{r}}\right)\Theta+\frac{\mu_{r}}{\mu_{d}+\mu_{r}}\dot{R}\frac{f''}{f'}\right]\dot{\del}^{k}_{r}\nn
\fl &+\frac{\mu_{d}}{1+w}\left[\left(\frac{w-c^{2}_{s}}{\mu_{d}+\mu_{r}}+\frac{4w}{3\mu_{d}+4\mu_{r}}-\frac{\mu_{r}}{3(\mu_{d}+\mu_{r})^{2}}\right)\dot{R}\Theta \frac{f''}{f'}+\left(\frac{4\mu_{r}}{3(3\mu_{d}+4\mu_{r})^{2}}\times\right.\right.\nn
\fl &\left.\left.\frac{4w\mu_{r}+(4w+1)\mu_{d}}{\mu_{d}+\mu_{r}}-\frac{4(c^{2}_{s}-w)}{3\mu_{d}+4\mu_{r}}-\frac{(c^{2}_{s}-w)c^{2}_{s}}{\mu_{d}+\mu_{r}}-\frac{c^{2}_{s}\mu_{r}}{3(\mu_{d}+\mu_{r})^{2}}\right)\Theta^{2}
\right.\nn
\fl &-\left.\left(\frac{1+w-2c^{2}_{s}}{\mu_{d}+\mu_{r}}+\frac{8w}{3\mu_{d}+4\mu_{r}}\right)\frac{\mu_{d}+\mu_{r}}{f'}+\left(\frac{4w}{3\mu_{d}+4\mu_{r}}-\frac{c^{2}_{s}}{\mu_{d}+\mu_{r}}\right)\frac{f}{f'}\right]\del^{k}_{d}\nn
\fl &+\frac{1}{1+w}\left[\left(\frac{(w-c^{2}_{s})\mu_{r}}{\mu_{d}+\mu_{r}}-\frac{3w\mu_{d}}{3\mu_{d}+4\mu_{r}}-\frac{\mu_{d}\mu_{r}}{3(\mu_{d}+\mu_{r})^{2}}\right)\dot{R}\Theta \frac{f''}{f'}
+\left(\frac{c^{2}_{s}\mu_{d}\mu_{r}}{3(\mu_{d}+\mu_{r})^{2}}\right.\right.\nn
\fl &\left.\left.+\frac{3(c^{2}_{s}-w)\mu_{d}}{3\mu_{d}+4\mu_{r}}-\frac{(c^{2}_{s}-w)c^{2}_{s}\mu_{r}}{\mu_{d}+\mu_{r}}+\frac{4\mu_{d}\mu_{r}}{(3\mu_{d}+4\mu_{r})^{2}}\frac{(1-3w)\mu_{r}-3w\mu_{d}}{3(\mu_{d}+\mu_{r})}\right)\Theta^{2}\right.\nn
\fl &\left.-\left(\frac{3w\mu_{d}}{3\mu_{d}+4\mu_{r}}+\frac{c^{2}_{s}\mu_{r}}{\mu_{d}+\mu_{r}}\right)\frac{f}{f'}
 -\left(\frac{\left(1+w-2c^{2}_{s}\right)\mu_{r}}{\mu_{d}+\mu_{r}}-
\frac{6w\mu_{d}}{3\mu_{d}+4\mu_{r}}\right)\times\right.\nn
\fl &\left.\frac{\mu_{d}+\mu_{r}}{f'}\right]\del^{k}_{r}+ \left[ \frac{1}{2}+\frac{k^{2}}{a^{2}}\frac{f''}{f'}-\frac{1}{2}\frac{ff''}{f'^{2}}+\frac{f''(\mu_{r}+\mu_{d})}{f'^{2}}
 -\dot{R}\Theta \left(\frac{f''}{f'}\right)^{2}\right.\nn
 \fl &\left.+\dot{R}\Theta\frac{f'''}{f'} \right] \car^{k}+\Theta \frac{f''}{f'}\dot{\car}^{k}=0\;.
 \end{eqnarray}
 \np
 In terms of the component perturbation variables  of section \ref{sec: variables} we can rewrite the propagation equation for the curvature fluid gradient as
 \begin{eqnarray}
\fl \ddot{\car}^{k}&+\left(2\dot{R}\frac{f'''}{f''}+\Theta\right)\dot{\car}^{k}+\left(\frac{k^{2}}{a^{2}}+\ddot{R}\frac{f'''}{f''}+\dot{R}^{2}\frac{f^{(iv)}}{f''}+\dot{R}\Theta \frac{f'''}{f''}+\frac{1}{3}\frac{f'}{f''}-\frac{R}{3}\right)\car^{k}\nonumber\\
\fl &+\left(\frac{c^{2}_{s}-1}{1+w}\frac{\mu_{d}}{\mu_{d}+\mu_{r}}-\frac{4wc^{2}_{z}}{1+w}\right)\dot{R}\dot{\del}^{k}_{d}+\left(\frac{c^{2}_{s}-1}{1+w}\frac{\mu_{r}}{\mu_{d}+\mu_{r}}+\frac{3wc^{2}_{z}}{1+w}\right)\dot{R}\dot{\del}^{k}_{r} \nonumber\\ 
\fl &+\left\{\left[(3c^{2}_{s}-1)\frac{\mu_{d}+\mu_{r}}{3f''}+\frac{w+c^{2}_{s}}{1+w}\dot{R}\Theta+\frac{c^{2}_{s}}{1+w}\left(2\ddot{R}+2\dot{R}^{2}\frac{f'''}{f''}\right)\right.\right.\nn
\fl &\left.\left.+\frac{\dot{R}}{1+w}\left(\dot{c^{2}_{s}}+c^{2}_{s}(c^{2}_{s}-w)\Theta\right)\right]\frac{\mu_{d}}{\mu_{d}+\mu_{r}}-4wc^{2}_{z}\left[\frac{\mu_{d}+\mu_{r}}{f''}+\frac{2}{1+w}\left(\ddot{R}+\dot{R}\Theta\right.\right.\right.\nn
\fl &\left.\left.\left.+\dot{R}^{2}\frac{f'''}{f''}\right)\right]+\frac{c^{2}_{s}-1}{3(1+w)}\frac{\mu_{d}\mu_{r}}{(\mu_{d}+\mu_{r})^{2}}\dot{R}\Theta\right\}\del^{k}_{d}+\left\{\left[(3c^{2}_{s}-1)\frac{\mu_{d}+\mu_{r}}{3f''}\right.\right.\nn
\fl &\left.\left.+\frac{w+c^{2}_{s}}{1+w}\dot{R}\Theta+\frac{c^{2}_{s}}{1+w}\left(2\ddot{R}+2\dot{R}^{2}\frac{f'''}{f''}\right)+\frac{\dot{R}}{1+w}\left(\dot{c^{2}_{s}}+c^{2}_{s}(c^{2}_{s}-w)\Theta\right)\right]\frac{\mu_{r}}{\mu_{d}+\mu_{r}}\right.\nn
\fl &\left.+3wc^{2}_{z}\left[\frac{\mu_{d}+\mu_{r}}{f''}+\frac{2}{1+w}\left(\ddot{R}+\dot{R}\Theta+\dot{R}^{2}\frac{f'''}{f''}\right)\right]\right.\nn
\fl &\left.-\frac{c^{2}_{s}-1}{3(1+w)}\frac{\mu_{d}\mu_{r}}{(\mu_{d}+\mu_{r})^{2}}\dot{R}\Theta\right\}\del^{k}_{r}=0\;.
 \end{eqnarray}
 %%%%%%%%%%%%%%%%%%%%%%%%%%%%%%%%%%%%%%%%%%%%%%%%%%%%
\section{Short Wavelength Solutions}
%%%%%%%%%%%%%%%%%%%%%%%%%%%%%%%%%%%%%%%%%%%%%%%%%%%%
In this section, we will study the evolution of the short wavelength modes, i.e., large values of  the wave number $k$, by using the equations presented in section 6 valid for a radiation and dust mixture. The  general results will  then be considered for 
$R^n$ models and a proposal for a {\it quasi-static} approximation for the matter perturbations
will be introduced for both radiation and dust dominated epochs. 
 In that approximation, widely used in the literature  \cite{QSA}, all the time 
derivative terms for the gravitational potentials are discarded, and only those
 including density perturbations are kept. The decoupling process for the involved
 equations is therefore highly simplified. Nonetheless, when this approximation was
used to study fourth order gravity theories in the metric formalism, it was proved
 as too aggressive and a more detailed analysis is required \cite{Dombriz_PRD, Aggressive}.

%%%%%%%%%%%%%%%%%%%%%%%%%%%%%%%%%%%%%%%%%%%%%%%%%%%%
\subsection{Perturbations in the Radiation-dominated Epoch}
%%%%%%%%%%%%%%%%%%%%%%%%%%%%%%%%%%%%%%%%%%%%%%%%%%%%
Let us now look at the case where the characteristic size of the fluid inhomogeneities is much less than the Jeans length for the radiation fluid but is still larger than the mean free path of the photon, i.e., 
$\lambda\ll \lambda_{H}\ll\lambda_{J}$. Similar investigation has been made by \cite{D1991} for the case of GR. 
Note that the scale dependence of the perturbations equations is not trivial
 as can be seen in \cite{ACD}.

Here we assume that we can neglect the interaction between the component fluids and that
 the radiation energy density can be taken as \textit{almost} homogeneous, meaning $\del_{r}\approx 0$.

This amounts to studying dust and curvature fluctuations on a homogeneous radiation background, whereby radiation affects the growth of the dust fluctuations by speeding up the cosmic expansion \cite{DBE}. We consider the flat ($K=0$) background, and hence the equations for such a background are given by
\begin{eqnarray}\label{totalf}
\fl \dot{\del}^{k}_{d}&+Z^{k}=\frac{\Theta}{h}\left(c^{2}_{s}\mu \del^{k}_{m}+p_{m} \vep^{k}\right)+a\left(\frac{k^{2}}{a^{2}}\right)V^{k}_{d}\;,\\
\fl \dot{Z}^{k}&-(\dot{R}\frac{f''}{f'}-\frac{2}{3}\Theta)Z^{k}-\left[ \frac{(2c^{2}_{s}-w-1)}{(1+w)}\frac{\mu_{m}}{f'} -\frac{c^{2}_{s}}{(1+w)}\left( \frac{R}{2}-\frac{f}{f'}-2\dot{R}\Theta\frac{f''}{f'}\right)\right]\del^{k}_{m}\nn
\fl &-\frac{w}{(1+w)}\left[2\frac{\mu_{m}}{f'}+ \frac{R}{2}-\frac{f}{f'}-2\dot{R}\Theta\frac{f''}{f'}\right] \vep^{k}
-\frac{1}{h}\left(\frac{k^{2}}{a^{2}}\right)\left(c^{2}_{s}\mu_{m}\del^{k}_{m}+p\vep^{k}\right)\nn
\fl &+\Theta\frac{f''}{f'}\Re^{k}-\left[ \frac{1}{2}+\frac{k^{2}}{a^{2}}\frac{f''}{f'}-\frac{1}{2}\frac{ff''}{f'^{2}}+\frac{f''\mu_{m}}{f'^{2}}-\dot{R}\Theta\left(\frac{f''}{f'}\right)^{2}+\dot{R}\Theta \frac{f'''}{f'}\right] {\car}^{k}=0\,,\\
\fl \dot{V}^{k}_{d}&+\frac{1}{3}\Theta V^{k}_{d}=\frac{1}{ah}\left(c^{2}_{s}\mu \del_{m}+p\varepsilon\right)\;,\\
\fl \dot{V}^{k}_{dr}&-\left(c^{2}_{z}-\frac{1}{3}\right)\Theta V^{k}_{dr}=\frac{1}{3ah}\mu\del^{k}_{m}-\frac{1}{a}c^{2}_{z}S^{k}_{dr}\;,\\
\fl \dot{\car}^{k}&=\Re^{k}-\frac{\dot{R}}{h}\left(c^{2}_{s}\mu_{m}\del^{k}_{m}+p_{m}\vep^{k}\right)\;,
\end{eqnarray}
\begin{eqnarray}
\fl \dot{\Re}^{k}&=-\left(2\dot{R}\frac{f'''}{f''}+\Theta\right)\Re^{k}-\dot{R}Z^{k}+\frac{\mu_{m}}{3f''}\del^{k}_{m}- \frac{1}{f''}\left(c^{2}_{s}\mu_{m}\del^{k}_{m}+p\vep^{k}\right)\nn
\fl &-\left(\frac{k^{2}}{a^{2}}+\ddot{R}\frac{f'''}{f''}+\dot{R}^{2}\frac{f^{(iv)}}{f''}+\dot{R}\Theta \frac{f'''}{f''}
+\frac{1}{3}\frac{f'}{f''}-\frac{R}{3}\right)\car^{k}.\label{totale}
\end{eqnarray}
Since $\del_{r}\ll\del_{d}$ we have
\be\label{entr}
c^{2}_{s}\mu \del^{k}_{m}+p \vep^{k}=\frac{1}{3}\mu_{r}\del^{k}_{r}\approx 0\;,
\ee
and so
\be\label{opy}
S^{k}_{dr}\approx \del^{k}_{d}\;.
\ee
In these limits the above set of equations (\ref{totalf}-\ref{totale}) can be rewritten as
\begin{eqnarray}\label{dotdel}
\fl &\dot{\del}^{k}_{d}+Z^{k}-a\left(\frac{k^{2}}{a^{2}}\right)V^{k}_{d}=0\;,\\
\fl &\dot{Z}^{k}-\left(\dot{R}\frac{f''}{f'}-2H\right)Z^{k}+\frac{\mu_{d}}{f'}\del^{k}_{d}
-\Theta\frac{f''}{f'}\Re^{k}\nn
\fl &~~-\left[ \frac{1}{2}+\frac{k^{2}}{a^{2}}\frac{f''}{f'}-\frac{1}{2}\frac{ff''}{f'^{2}}+\frac{f''\mu_{r}}{f'^{2}}-3H\dot{R}\left(\frac{f''}{f'}\right)^{2}+3H\dot{R}\frac{f'''}{f'}\right] {\car}^{k}=0\;,\\
\fl &\dot{V}^{k}_{d}+HV^{k}_{d}=0\;,
\end{eqnarray}
\begin{eqnarray}
\fl &\dot{V}^{k}_{dr}+\frac{4}{3}\frac{\mu_{r}}{h}H V^{k}_{dr}=0\;,\\
\fl &\dot{\car}^{k}= \Re^{k},\\ \label{dotre}
\fl &\dot{\Re}^{k}=-\left(2\dot{R}\frac{f'''}{f''}+3H\right)\Re^{k}-\dot{R}Z^{k}+\frac{\mu_{d}}{3f''}\del^{k}_{d}\nn
\fl &~~~~~~-\left(\frac{k^{2}}{a^{2}}+\ddot{R}\frac{f'''}{f''}+\dot{R}^{2}\frac{f^{(iv)}}{f''}+3H\dot{R}\frac{f'''}{f''}
+\frac{1}{3}\frac{f'}{f''}-\frac{R}{3}\right)\car^{k}\;.
\end{eqnarray}
From Eqns. (\ref{dotdel})-(\ref{dotre}) we obtain  the following two second order differential equations: 
\begin{eqnarray}
\fl &\ddot{\del}^{k}_{d}+\left(2H-\frac{3\dot{R}f''}{4f'}\frac{\mu_{d}}{\mu_{r}}\right)\dot{\del}^{k}_{d}-\frac{\mu_{d}}{f'}\del^{k}_{d}
+3H\frac{f''}{f'}\dot{\car}^{k}\nn
\fl &~~+\left[ \frac{1}{2}+\frac{k^{2}}{a^{2}}\frac{f''}{f'}-\frac{1}{2}\frac{ff''}{f'^{2}}+\frac{f''\mu_{r}}{f'^{2}}-3H\dot{R}\left(\frac{f''}{f'}\right)^{2}+3H\dot{R}\frac{f'''}{f'}\right] {\car}^{k}=0\;,\\
\fl &\ddot{\car}^{k}+\left(2\dot{R}\frac{f'''}{f''}+3H\right)\dot{\car}^{k}-\frac{3\dot{R}\mu_{d}}{4\mu_{r}}\dot{\del}^{k}_{d}-\frac{\mu_{d}}{3f''}\del^{k}_{d}\nn
\fl &~~~~+\left(\frac{k^{2}}{a^{2}}+\ddot{R}\frac{f'''}{f''}+\dot{R}^{2}\frac{f^{(iv)}}{f''}+3H\dot{R}\frac{f'''}{f''}
+\frac{f'}{3f''}-\frac{R}{3}\right)\car^{k}=0\;.
\end{eqnarray}
It can be shown that $H$ and $\dot{R} f''/f'$  are of the same behaviour for $R^{n}$ models, whereas $\mu_{d}\ll \mu_{r}$, implying that curvature and radiation fluids effectively dominate the fluctuation dynamics. In effect, terms like $\mu_{d}\del^{k}_{d}$ merely sub-dominate in the curvature-radiation-dust mixture. Hence we can safely approximate the above equations by
\begin{eqnarray}\label{ddotdr}
\fl &\ddot{\del}^{k}_{d}+2H\dot{\del}^{k}_{d}+3H\frac{f''}{f'}\dot{\car}^{k}
+\left[ \frac{1}{2}\left(1-\frac{ff''}{f'^{2}}\right)+\frac{k^{2}}{a^{2}}\frac{f''}{f'}+\frac{f''\mu_{r}}{f'^{2}}-3H\dot{R}\left(\frac{f''}{f'}\right)^{2}\right.\nn
\fl &\left.+3H\dot{R}\frac{f'''}{f'}\right] {\car}^{k}=0\;,
\label{ddotrr}\\
\fl &\ddot{\car}^{k}+\left(2\dot{R}\frac{f'''}{f''}+3H\right)\dot{\car}^{k}
+\left(\frac{k^{2}}{a^{2}}+\ddot{R}\frac{f'''}{f''}+\dot{R}^{2}\frac{f^{(iv)}}{f''}+3H\dot{R}\frac{f'''}{f''}+\frac{f'}{3f''}-\frac{R}{3}\right)\car^{k}=0\;.\nn
\fl&
\end{eqnarray}
These two equations tell us that, deep in the radiation-dominated era, the curvature fluctuations are decoupled from the matter in the second order equations.

GR is a specific example of the generalized $R^{n}$ models where $n=1$. In this limit, Eqns. (\ref{ddotdr}) and (\ref{ddotrr})  reduce to
\begin{eqnarray}
&\ddot{\del}^{k}_{d}+2H\dot{\del}^{k}_{d}+\frac{1}{2} {\car}^{k}=0\;,\\
&\car^{k}=0\;,
\end{eqnarray}
thus yielding the standard  GR equation  for the density contrast in a radiation background
\be
\ddot{\del}^{k}_{d}+\frac{1}{t}\dot{\del}^{k}_{d}=0\;,
\ee
whose general solution is given by
\be
\del^{k}_{d}(t)=C_{1}+C_{2}\ln{t}\;.
\ee
with $C_{1,2}$ arbitrary constants. For $n\neq 1$, with $w=\frac{1}{3}$ in the radiation-dominated epoch,  Eqns. (\ref{ddotdr}) and (\ref{ddotrr}) take the following forms:
\begin{eqnarray}
\label{reqd}
&\ddot{\del}^{k}_{d}+\frac{n}{t}\dot{\del}^{k}_{d}
+\frac{t}{2}\dot{\car}^{k}+\left[\frac{12-5n}{4}+\frac{n}{12}\left(\frac{\lambda_{H}}{\lambda}\right)^{2}_{eq}t^{2-n}\right] {\car}^{k}=0\;,\\ 
\label{reqre}
&\ddot{\car}^{k}-\left(\frac{5n-16}{2t}\right)\dot{\car}^{k}+\left[\frac{n^{2}}{4}\left(\frac{\lambda_{H}}{\lambda}\right)^{2}_{eq}t^{-n}-\frac{6(n-2)}{t^{2}}\right]\car^{k}=0\;,
\end{eqnarray}
where we have used the fact that  $\frac{k^{2}}{a^{2}}=\frac{n^{2}}{4}\left(\frac{\lambda_{H}}{\lambda}\right)^{2}_{eq}t^{-n}$ 
with normalized time $t_{eq}=1$ at the time of  \textit{radiation-matter equality}.
%%%%%%%%%%%%%%%%%%%%%%%%%%%%%%%%%%%%%%%%%%%%%%%%%%%%
\subsubsection*{Quasi-static Analysis}
%%%%%%%%%%%%%%%%%%%%%%%%%%%%%%%%%%%%%%%%%%%%%%%%%%%%
In general the system of equations (\ref{reqd})-(\ref{reqre}) yield Bessel hypergeometric type  analytic solutions. However, since we are dealing with small scales we can take a   {\it quasi-static approximation } where the time variations in $\car^{k}$ are treated as negligible, i.e., $\ddot{\car}^{k}\simeq 0$ and $\dot{\car}^{k}\simeq 0$. In this scheme the overall dynamics of the density perturbations leads to  the  simplified, $k$- independent, equation
\be
\ddot{\del}^{k}_{d}+\frac{n}{t}\dot{\del}^{k}_{d}=0\;.
\label{eq_rad_QSA}
\ee
This equation admits the general solution
\be
\del^{k}_{d}(t)=C_{1}+C_{2}t^{1-n}\;.
\ee
 
On  small scales, radiation suppresses the growth of fluctuations as they enter the horizon before radiation-dust equality, and 
dust (baryon) self-gravitation is not yet strong enough to offset the cosmic expansion. This is because 
the expansion scale factor grows faster than the perturbation amplitudes do. The phenomenon is known 
in the literature as the {\it M\'esz\'aros effect} \cite{P1974}.

It is clear from the above analysis that the  M\'esz\'aros effect puts a constraint on the value of $n$ in $R^{n}$ gravity. To do so, all we need do is  determine the allowed values of $n$ for which the perturbation amplitudes grow slower than the expansion in the radiation dominated era, i.e.,
\begin{equation}
\frac{d}{dt}\left[\frac{\del^{k}_{d}(t)}{a(t)}\right]\propto \frac{d}{dt}\left[\frac{t^{1-n}}{t^{\frac{n}{2}}}\right]<0\Rightarrow 1-\frac{3n}{2} <0\;.
\end{equation}
This means that only values of $n>\frac{2}{3}$ give a growth rate compatible with the M\'esz\'aros effect.

In figure 2, we plot the normalized dust density contrast $\delta(t)\equiv \Delta_{m}(t)/\Delta_{eq}$ in the radiation-dominated epoch. 
%The figure on the right shows the relative error of the full solutions and the quasi-static approximation.
\begin{figure*}[h!]
	\centering
		\includegraphics[width=1.0\textwidth]{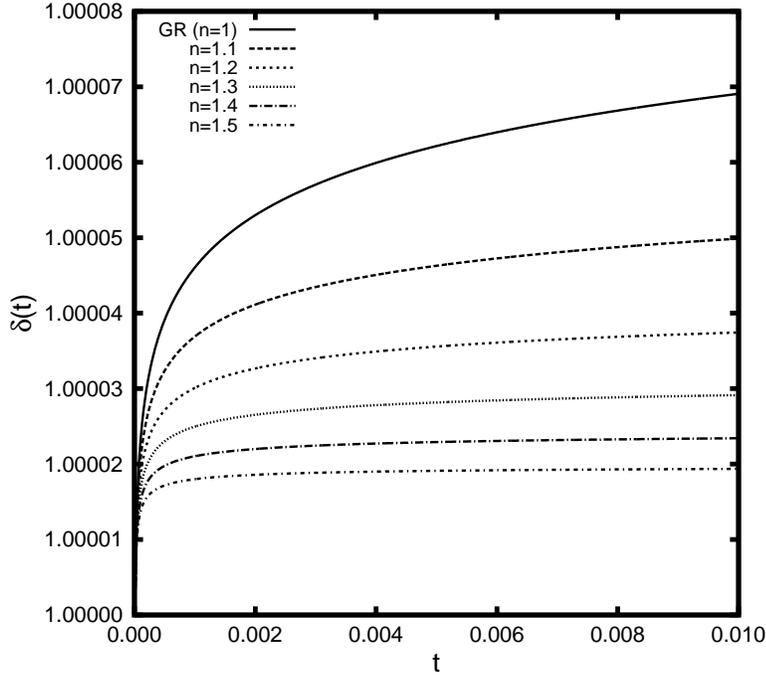}\,\,\,\,\,\,\,\,\,\,\,\,\,\,
		\caption{\footnotesize{
		Dust growth factor in the radiation dominated epoch for $R^n$ models: The plots show the growth factor obtained by solving numerically the full system of  equations (\ref{reqd}) and (\ref{reqre}) for scale $\lambda_{H}=100\lambda$ and the quasi-static solution (\ref{eq_rad_QSA}) for $n=1.5,1.4,1.3,1.2,1.1$ from bottom to top. The top-most plot corresponds to GR ($n=1$). It can be seen that quasi-static results are quite close to those of the full system for the stated values of $n$, only slightly (but invisibly) lower in the plots. For values of $\lambda_{H}>100\lambda$ the growth factor appears to be insensitive to scale showing the convenience of introducing the {\it quasi-static} approximation. 
% \label{reqd} 133
%\label{reqre} 134
%\label{eq_rad_QSA} 
		}}
	%\label{fig:rangoespin}
\end{figure*}
%%%%%%%%%%%%%%%%%%%%%%%%%%%%%%%%%%%%%%%%%%%%%%%%%%%%
\subsection{Perturbations in the Dust-dominated Epoch}
%%%%%%%%%%%%%%%%%%%%%%%%%%%%%%%%%%%%%%%%%%%%%%%%%%%%
During this epoch of the universe  the dust energy  density  is dominating in the two-fluid dynamics and all order-of-magnitude approximations go in line with the assumption that $\mu_{d}>>\mu_{r}$. The equations (\ref{totalf})-(\ref{totale}) will,
upon imposing the short-wavelength assumptions (\ref{entr},\ref{opy}), become
\begin{eqnarray}
\fl &\dot{\del}^{k}_{d}+Z^{k}+a\tl\nb^{2}V^{k}_{d}=0\;,\\
\fl &\dot{Z}^{k}-\left(\dot{R}\frac{f''}{f'}-2H\right)Z^{k}+\frac{\mu_{d}}{f'}\del^{k}_{d}
-\Theta\frac{f''}{f'}\Re^{k}\nn
\fl &~~-\left[ \frac{1}{2}+\frac{k^{2}}{a^{2}}\frac{f''}{f'}-\frac{1}{2}\frac{ff''}{f'^{2}}+\frac{f''\mu_{d}}{f'^{2}}-3H\dot{R}\left(\frac{f''}{f'}\right)^{2}+3H\dot{R}\frac{f'''}{f'}\right] {\car}^{k}=0\;,\\
\fl &\dot{V}^{k}_{d}+HV^{k}_{d}=0\;,\\
\fl &\dot{V}^{k}_{dr}+\frac{4}{3}\frac{\mu_{r}}{h}HV^{k}_{dr}=0\Rightarrow \dot{V}^{k}_{dr}+\frac{4\mu_{r}}{3\mu_{d}}HV^{k}_{dr}=0\;,\\
\fl &\dot{\car}^{k}= \Re^{k}\;,
\end{eqnarray}
\begin{eqnarray}
\fl &\dot{\Re}^{k}=-\left(2\dot{R}\frac{f'''}{f''}+3H\right)\Re^{k}-\dot{R}Z^{k}+\frac{\mu_{d}}{3f''}\del^{k}_{d}\nn
\fl &~~~~~-\left[\frac{k^{2}}{a^{2}}+(3H\dot{R}+\ddot{R})\frac{f'''}{f''}+\dot{R}^{2}\frac{f^{(iv)}}{f''}
+\frac{f'}{3f''}-\frac{R}{3}\right]\car^{k}\;.
\end{eqnarray}
The resulting  set of second order equations is therefore
\begin{eqnarray}
\fl &\ddot{\del}^{k}_{d}+\left(2H-\dot{R}\frac{f''}{f'}\right)\dot{\del}^{k}_{d}-\frac{\mu_{d}}{f'}\del^{k}_{d}
+3H\frac{f''}{f'}\dot{\car}^{k}\nn
\fl &~~~~+\left[ \frac{1}{2}+\frac{k^{2}}{a^{2}}\frac{f''}{f'}-\frac{ff''}{2f'^{2}}+\frac{f''\mu_{d}}{f'^{2}}-3H\dot{R}\left(\frac{f''}{f'}\right)^{2}+3H\dot{R}\frac{f'''}{f'}\right] {\car}^{k}=0\;,  \label{ddotdd} \\ 
\fl &\ddot{\car}^{k}+\left(2\dot{R}\frac{f'''}{f''}+3H\right)\dot{\car}^{k}-\dot{R}\dot{\del}^{k}_{d}-\frac{\mu_{d}}{3f''}\del^{k}_{d}\nn
\fl &~~+\left[\frac{k^{2}}{a^{2}}+(3H\dot{R}+\ddot{R})\frac{f'''}{f''}+\dot{R}^{2}\frac{f^{(iv)}}{f''} +\frac{f'}{3f''}-\frac{R}{3}\right]\car^{k}=0\;.
\label{ddotrd}
\end{eqnarray}
As can be observed, these two equations differ from their counterparts in the  radiation-dominated epoch in that the curvature perturbations are \textit {not} decoupled from 
 that of matter  in the system of equations.

The limiting GR perturbation equations for (\ref{ddotdd}) and (\ref{ddotrd}) in this epoch are given by
\begin{eqnarray}
&\ddot{\del}^{k}_{d}+2H\dot{\del}^{k}_{d}-\mu_{d}\del^{k}_{d}
+ \frac{1}{2} {\car}^{k}=0\;,\\ 
&~~-\frac{\mu_{d}}{3}\del^{k}_{d}+\frac{1}{3}\car^{k}=0\;,
\end{eqnarray} 
and combine to give the equation
\be
\ddot{\del}^{k}_{d}+\frac{4}{3t}\dot{\del}^{k}_{d}-\frac{2}{3t^{2}}\del^{k}_{d}=0\;.
\ee
This equation admits the well known solution
\be
\del^{k}_{d}(t)=C_{1}t^{-1}+C_{2}t^{\frac{2}{3}}\;.
\ee
For $R^{n}$ models, equations (\ref{ddotdd},\ref{ddotrd}) take the form

\begin{eqnarray}\label{mesdrr}
\fl &\ddot{\del}^{k}_{d}+\left(\frac{10n-6}{3t}\right)\dot{\del}^{k}_{d}+\frac{2(8n^{2}-13n+3)}{3t^{2}}\del^{k}_{d}
+\frac{3(n-1)}{2(4n-3)}t\dot{\car}^{k}\nn
\fl &~~~~+\left[\frac{n(n-1)}{3(4n-3)}\left(\frac{\lambda_{H}}{\lambda}\right)^{2}_{eq}t^{2-\frac{4n}{3}}+\frac{27n^{2}-8n^{3}-18n}{2n(4n-3)}\right] {\car}^{k}=0\;,\\
\label{mesrrr}
\fl &\ddot{\car}^{k}+\left\{\frac{8n\left[n(8n-13)+3\right](4n-3)}{27(n-1)t^{4}}\right\}\del^{k}_{d}+\frac{8n(4n-3)}{3t^{3}}\dot{\del}^{k}_{d}+\frac{8-2n}{t}\dot{\car}^{k}\nn
\fl &~~~~+\left\{\frac{4n^{2}}{9}\left(\frac{\lambda_{H}}{\lambda}\right)^{2}_{eq}t^{-\frac{4n}{3}}-\frac{2\left[n(8n+5)-69\right]+54}{9(n-1)t^{2}}\right\}\car^{k}=0\;,
\end{eqnarray}
where $\frac{k^{2}}{a^{2}}=\frac{4n^{2}}{9}\left(\frac{\lambda_{H}}{\lambda}\right)^{2}_{eq}t^{-\frac{4n}{3}}$ during this epoch.
%%%%%%%%%%%%%%%%%%%%%%%%%%%%%%%%%%%%%%%%%%%%%%%%%%%%
\subsubsection*{ Quasi-static Analysis}
%%%%%%%%%%%%%%%%%%%%%%%%%%%%%%%%%%%%%%%%%%%%%%%%%%%%
In the quasi-static limit with $\left(\frac{\lambda_{H}}{\lambda}\right)^{2}_{eq}>>1$ we get a single second order $k$-scale independent  equation 
\be
\ddot{\del}^{k}_{d}+\frac{4n}{3t}\dot{\del}^{k}_{d}+\left[\frac{4(8n^{2}-13n+3)}{9t^{2}}\right]\del^{k}_{d}=0\;,
\label{dust_dust_QSA}
\ee
the solution of which is given by
\be\label{sol}
\del^{k}_{d}(t)=C_{1}t^{\alpha_{+}}+C_{2}t^{\alpha_{-}}\;,
\ee
where  $\alpha_{\pm}=-\frac{2n}{3}+\frac{1}{2}\pm\frac{\sqrt{-112n^{2}+184n-39}}{6}$. 
The coefficients $C_{1,2}$ can be determined by imposing initial conditions.

At $t=t_{eq}=1$ we have 
\be\label{ctandco}
\del^{k}_{(d)\,eq} \equiv \del^{k}_{(d)}(t_{eq})=C_{1}+C_{2}\;,
\ee
and differentiating (\ref{sol}) gives
\be
\dot{\del}^{k}_{d}(t)=\alpha_{+}C_{1}t^{\alpha_{+}-1}+\alpha_{-}C_{2}t^{\alpha_{-}-1}\;,
\ee
which, at equality, will give
\be\label{coanct}
\dot{\del}^{k}_{(d)\,eq} \equiv \dot{\del}^{k}_{(d)}(t_{eq})=\alpha_{+}C_1+\alpha_{-}C_2\;.
\ee
Solving (\ref{ctandco}) and (\ref{coanct}) simultaneously we obtain
\be
C_{1,2}=\frac{\pm\dot{\Delta}^{k}_{(d)\,eq}\mp\alpha_{\mp}\Delta^{k}_{(d)\,eq}}{\alpha_{+}-\alpha_{-}} \;.
\ee
Fig. 3  shows the evolution of the density perturbations $\delta(t)\equiv \del^{k}_{(d)}(t)/\del^{k}_{(d)eq}$ in time ($t$ from 1 onwards, where $t=t_{eq}=1$ is the normalized time at equality) for the above linearly independent solutions, $C_{1,2}$ having been obtained by setting $\del^{k}_{(d)eq}=10^{-5}$ and $\dot{\del}^{k}_{(d)eq}=10^{-5}$ .
 \begin{figure*}[h!]
	\centering
		\includegraphics[width=1.0\textwidth]{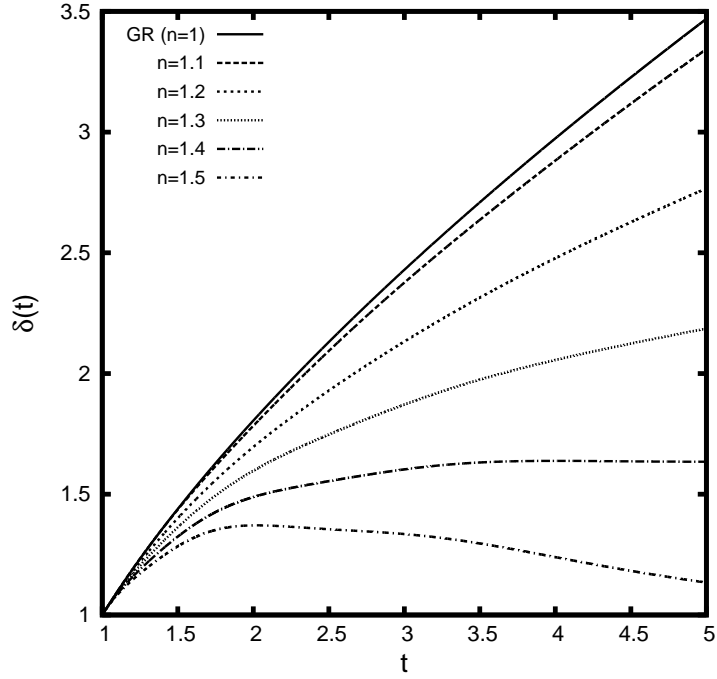}\,\,\,\,\,\,\,\,\,\,\,\,\,\,
		\caption{\footnotesize{Dust growth factor in the dust dominated epoch for $R^n$ models: The plots show the growth factor obtained by solving numerically the full system equations (\ref{mesdrr}) and (\ref{mesrrr}) for scale $\lambda_{H}=100\lambda$ and  the quasi-static analytic solution (\ref{dust_dust_QSA}) . It can be seen that quasi-static results are indistinguishable from the full results for  $n=1.5, 1.4, 1.3,1.2, 1.1$ from bottom to top, with the full system solutions slightly higher than those of the quasi-static approximation. It can also be seen that for higher values of $n$ the growth factor increases more slowly till a critical value of $n$ somewhere between 1.4 \& 1.5 where the growth factor becomes a decreasing function of time. Note the $n=1$ case (GR) is presented on the topmost plot.
% \label{mesdrr} 150 
%\label{mesrrr} 151
%\label{dust_dust_QSA} 152		
		}}
	\label{fig:rangoespin2}
\end{figure*}
%%%%%%%%%%%%%%%%%%%%%%%%%%%%%%%%%%%%%%%%%%%%%%%%%%%%
\section{Long Wavelength Solutions}
%%%%%%%%%%%%%%%%%%%%%%%%%%%%%%%%%%%%%%%%%%%%%%%%%%%%
For specific intervals of $n$, a set of initial conditions give rise to cosmic histories which include a transient decelerated phase which evolves towards an accelerated phase. Structure formation takes place during the transient regime \cite{CDT}.

In this section we analyze the evolution of scalar perturbations during this phase, in the \textit{long wavelength limit}. In this limit the wavenumber $k$ is so small that $\lambda=\frac{2\pi a}{k}\gg \lambda_{H}$, i.e.,  $\frac{k^{2}}{a^{2}H^{2}}\ll1$.
All Laplacian terms  can therefore be neglected and
 spatially flat $(K=0)$ backgrounds guarantee the conservation of $C$, i.e., $\dot{C}^{k}=0$.  
In this paper  we are considering 
only adiabatic perturbations, i.e. $S_{ij}=0$ and hence, for a radiation-dust mixture, the equation for the evolution of entropy perturbations
\be
\dot{S}_{dr}+a\tilde{\nabla}^{2}V_{dr}=0\;.
\ee
implies that 
\be V_{dr}=0\;.\ee
And from this and the equation
\be
\dot{V}_{dr}-\left(c^{2}_{z}-\frac{1}{3}\right)\Theta V_{dr}=-\frac{1}{ah}(c^{2}_{sd}-c^{2}_{sr})\mu\del_{m}-\frac{1}{a}c^{2}_{z}S_{dr}\;.
\ee
follows
\be
(c^{2}_{sd}-c^{2}_{sr})\mu\del_{m}=0\;.
\ee
We therefore have the following system of equations:
\begin{eqnarray}
\fl &\dot{\del}_{m}+(1+w)Z-w\Theta\del_{m}=0\;,\\
\fl &\dot{Z}-\left(\dot{R}\frac{f''}{f'}-\frac{2}{3}\Theta\right)Z-\left[ \frac{(2c^{2}_{s}-w-1)}{(1+w)}\frac{\mu_{m}}{f'} +\frac{c^{2}_{s}}{(1+w)}\left( \frac{R}{2}-\frac{f}{f'}-2\dot{R}\Theta\frac{f''}{f'}\right)\right]\del_{m}\nn
\fl &~~~-\Theta\frac{f''}{f'}\Re-\left[ \frac{1}{2}-\frac{1}{2}\frac{ff''}{f'^{2}}+\frac{f''\mu_{m}}{f'^{2}}-\dot{R}\Theta\left(\frac{f''}{f'}\right)^{2}+\dot{R}\Theta \frac{f'''}{f'}\right] {\car}=0\;,\\
\fl &\dot{\car}-\Re+\frac{c^{2}_{s}}{1+w}\dot{R}\del_{m}=0\;,
\end{eqnarray}
\begin{eqnarray}
\fl &\dot{\Re}+\left(2\dot{R}\frac{f'''}{f''}+\Theta\right)\Re+\dot{R}Z+ \frac{(3c^{2}_{s}-1)\mu_{m}}{3f''}\del_{m} %\nn
\nn
\fl &+\left[(\dot{R}\Theta+\ddot{R})\frac{f'''}{f''}+\dot{R}^{2}\frac{f^{(iv)}}{f''}
+\frac{f'}{3f''}-\frac{R}{3}\right]\car=0\;,\nn
\fl &\frac{C_{0}}{a^{2}}+\left(\frac{4}{3}\Theta +2\frac{\dot{R}f''}{f'} \right) Z-2\frac{\mu_{m}}{f'}\del_{m}+\left[2\dot{R}\Theta \frac{f'''}{f'}-\frac{f''}{f'}\left(\frac{f}{f'}-2\frac{\mu_{m}}{f'}+2\dot{R}\Theta\frac{f''}{f'}\right)\right] \car\nn
\fl &~~~~~+ 2\Theta\frac{ f''}{f'}\Re=0,\label{C}
\end{eqnarray}
$C_{0}$ being the conserved value for the quantity $C$.

In terms of the background $R^{n}$ solutions and making use of the conservation of $C$ the above equations can be rewritten as
\begin{eqnarray}
\fl \dot{\del}_{m}&=\left[\frac{1+w-2n}{1+w}-\frac{6(n-1)n}{n+3(n-1)w-3}\right]\frac{\del_{m}}{t}\nn
\fl &-\frac{3(1+w)^{2}}{4a^{2}_{0}\left[n+3(n-1)w-3\right]\left[4n-3(1+w)\right]}t^{1-\frac{4n}{3(1+w)}}C_{0}\nn
\fl &- \frac{ 9(n-1)(1+w)^{3} t^{2}}{4\left[n+3(n-1)w-3\right]\left[4n-3(1+w)\right]}t^{2}\Re
\nn
\fl & +\left\{\frac{3(n-1)(1+w)^{2}\left[n(6w+8)-15(1+w)\right]}{4\left[n+3(n-1)w-3\right]\left[4n-3(1+w)\right]}\right\} t\car\;,\nn
\fl
\end{eqnarray}
\begin{eqnarray}
\fl \dot{\car}= \Re+\frac{8nc^{2}_{s}\left[4n-3(1+w)\right]}{3(1+w)^{3}}\frac{\del_{m}}{t^{3}}\;,
\end{eqnarray}
\begin{eqnarray}
\fl \dot{\Re}&=-2\left[\frac{(n-4)+2(n-2)w}{(1+w)t}-\frac{3n(n-1)}{n+3w(n-1)-3}\right]\Re\nn
\fl &+\frac{2n(4n-3w-3)}{(1+w)\left[n+3(n-1)w-3\right]}C_{0}t^{-\frac{4n}{3(1+w)}-2}\nn
\fl &-2\left[\frac{9n(n-2)(n-1)}{n+3(n-1)w-3}+2n^{2}-7n-\frac{3n^{2}(9n-26)+57n}{9(1+w)(n-1)}-\frac{8n^{2}(n-2)}{9(1+w)^{2}(n-1)}\right.\nn
\fl &\left.+6\right]\frac{\car}{t^{2}} +16n\frac{\del_{m}}{t^{4}}
\frac{\left[4n-3(1+w)\right]\left[4n+3(n-1)w-3\right]}{27(n-1)(1+w)^{4}\left[n+3(n-1)w-3\right]}\times\nn
\fl &  \left[\left(9w(1+w)+8\right)n^{2}-\left(27w^2+24w+13\right)n+3(1+w)(1+6w)\right]\;.
\end{eqnarray}
%%%%%%%%%%%%%%%%%%%%%%%%%%%%%%%%%%%%%%%%%%%%%%%%%%%%
\subsection{Perturbations in the Radiation-dominated Epoch}
%%%%%%%%%%%%%%%%%%%%%%%%%%%%%%%%%%%%%%%%%%%%%%%%%%%%
The second order set of equations governing the dynamics of density perturbations in this epoch is given by

\begin{eqnarray}
\fl &\ddot{\del}^{k}_{r}+\frac{n(9n-14)+4}{2(n-2)t}\dot{\del}^{k}_{r}+\frac{n\left[n(n(19n-54)+58)-32\right]+8}{2(n-2)^{2}t^{2}}\del^{k}_{r}\nn
\fl &~~+\frac{2\left[n(3n-4)+2\right]}{3(n-2)^{2}}t\dot{\car^{k}}-\frac{n(15n-22)+14}{3(n-2)}{\car^{k}}+\frac{4(n^{2}-1)}{3(n-2)^{2}}t^{-n}C_{0}=0,\\
\fl &\ddot{\car}^{k}-\frac{n(11n-32)+32}{2(n-2)t}\dot{\car}^{k}+\frac{3\left[n(5n-9)+8\right]}{2t^{2}}{\car^{k}}-\frac{3n\left[n(n-3)+2\right]}{2(n-2)t^{3}}\dot{\del}^{k}_{r}\nn
\fl &~~-\frac{3n(n-1)\left[n(19n-28)+4\right]}{4(n-2)t^{4}}\del^{k}_{r}-\frac{3n(n-1)}{(n-2)}t^{-(n+2)}C_{0}=0\;.
\end{eqnarray}
Making use of the conservation of $C$, we can eliminate $\dot{\car}^{k}$ and 
$\car^{k}$ quantities in favour of $\del^{k}_{r}$ (and its derivatives) and $C_{o}$. This way we can get a decoupled third order k-scale independent equation for $\del^{k}_{r}$:
\begin{eqnarray}
\fl \frac{d^3}{dt^3}\del^{k}_{r}&-\frac{n-5}{t}\frac{d^2}{dt^2}\del^{k}_{r}+\frac{24n-19n^2+8}{4t^{2}}\frac{d}{dt}\del^{k}_{r}+\frac{(n-2)\left[5n^2-8n+2\right]}{2t^{3}}\del^{k}_{r}-\frac{(12-7n)C_{0}}{3t^{(n+1)}}=0\;.\nn
\fl&
\end{eqnarray}
This equation admits the general solution 
\be
\del^{k}_{r}(t)=C_{1}t^{\frac{n}{2}-1}+C_{2}t^{\beta_{+}}+C_{3}t^{\beta_{-}}+C_{4}t^{2-n}\;.
\ee
where $C_{1,2,3}$ are arbitrary integration constants to be evaluated from initial conditions with
\be
C_{4}\equiv \frac{2\left(24-14n\right)C_{0}}{9(7n^{3}-18n^{2}+16)}
\ee
and 
\be
\beta_{\pm} \equiv-\frac{1}{2}+\frac{n}{4}\pm\frac{\sqrt{3(81n^{2}-44n+12)}}{4}\;.
\ee
Provided that the initial values of  $\Delta^{k}_{r}$, $\dot{\Delta}^{k}_{r}$, $\ddot{\Delta}^{k}_{r}$ and $C_{0}$ are known at $t_{eq}=1$, 
the integration constants can be determined since 
%from (\ref{intcont}) 
%provided that 
\begin{eqnarray}\label{intcont}
\fl\Delta^{k}_{(r)eq}&=C_{1}+C_{2}+C_{3}+C_{4},\nn
\fl\dot{\Delta}^{k}_{(r)eq}&=\left(\frac{n-2}{2}\right)C_{1}+C_{2}\beta_{+}+C_{3}\beta_{-}+(2-n)C_{4}\;,\nn
\fl\ddot{\Delta}^{k}_{(r)eq}&=\left[\frac{(n-2)(n-4)}{4}\right]C_{1}+C_{2}\beta_{+}(\beta_{+}-1)\nn
\fl &+C_{3}\beta_{-}(\beta_{-}-1)+(2-n)(1-n)C_{4}\;.
\end{eqnarray}
We do not present $C_{1,2,3}$ explicitly for the sake of simplicity.
%%%%%%%%%%%%%%%%%%%%%%%%%%%%%%%%%%%%%%%%%%%%%%%%%%%%
\subsection{Perturbations in the Dust-dominated Epoch }
%%%%%%%%%%%%%%%%%%%%%%%%%%%%%%%%%%%%%%%%%%%%%%%%%%%%
Proceeding in a similar fashion for the  dust dominated, long wavelength, regime gives the second order evolution equations given by
\begin{eqnarray}
\fl \ddot{\del}^{k}_{d}&+\frac{n(8n-13)+3}{(n-3)t}\dot{\del}^{k}_{d}+\frac{\left[n(8n-13)+3\right]\left[n(16n-15)+9\right]}{3(n-3)^{2}t^{2}}\del^{k}_{d}\nn
\fl &+\frac{3(n-1)\left[n(16n-15)+9\right]}{4(n-3)^{2}(4n-3)}t\dot{\car^{k}}
-\frac{n\left[\left(n(16n(8n-31)+711\right)-540\right]+189}{4(n-3)^{2}(4n-3)}{\car^{k}}\nn
\fl &-\frac{n\left(27+54n-56n^{2}\right)-27}{4(n-3)^{2}(4n-3)}t^{-\frac{4n}{3}}C_{0}=0\;,\\
\fl &\ddot{\car}^{k}-\frac{4(n-1)\left[n(2n-5)+6\right]}{\left[n(n-4)+3\right)t}\dot{\car}^{k}+\frac{4\left[n\left(n(2n(16n-65)+213)-198\right)+81\right]}{9\left[n(n-4)+3\right]t^{2}}{\car^{k}}\nn
\fl &-\frac{16n(3-4n)^{2}\left[n(8n-13)+3\right]}{27\left[(n-4)n+3\right]t^{4}}\del^{k}_{d}-\frac{2n\left[n(4n-7)+3\right]}{n(n-4)+3}t^{-(n+2)}C_{0}=0\;,
\end{eqnarray}
which reduce to a single third order evolution equation for the density perturbations given as
\begin{eqnarray}
\label{sold}
\fl \frac{d^3}{dt^3}\del^{k}_{d}&+\frac{5}{t}\frac{d^2}{dt^2}\del^{k}_{d}-\frac{2\left[n\left(4n(8n-19)+33\right)+9\right]}{9(n-1)t^{2}}\frac{d}{dt}\del^{k}_{d}-\frac{2(4n-3)\left[n(8n-13)+3\right]}{9(n-1)t^{3}}\del^{k}_{d}\nn
\fl &-\frac{\left(12n^2-31n+18\right)}{6(n-1)\,t^{n+1}}C_{0}=0\;,
\end{eqnarray}
which is a third order decoupled k-scale independent equation.  The general solution of (\ref{sold}) is given by
\be
\del^{k}_{d}(t)=C_{1}t^{-1}+C_{2}t^{\gamma_{+}}+C_{3}t^{\gamma_{-}}+C_{4}t^{2-\frac{4n}{3}}\;,
\ee
where $C_{1,2,3}$ are arbitrary integration constants to be evaluated from initial conditions and 
\be
C_{4}\equiv\frac{9\left(12n^{2}-31n+18\right)C_{0}}{8(48n^{4}-184n^{3}+159n^{2}+63n-81)}
\ee
together with
\be
\gamma_{\pm} \equiv-\frac{1}{2}\mp\frac{1}{6}\sqrt{\frac{256n^{3}-608n^{2}+417n-81}{n-1}}\;.
\ee

As in the radiation epoch, the integration constants $C_{1,2,3}$ can be determined  from 
the initial values of  $\Delta^{k}_{d}$, $\dot{\Delta}^{k}_{d}$, $\ddot{\Delta}^{k}_{d}$  and $C_{0}$ known at $t_{eq}=1$ as follows:
\begin{eqnarray}\label{intcont2}
&\Delta^{k}_{(d)eq}=C_{1}+C_{2}+C_{3}+C_{4},\nn
&\dot{\Delta}^{k}_{(d)eq}=-C_{1}+C_{2}\gamma_{+}+C_{3}\gamma_{-}+\left(\frac{6-4n}{3}\right)C_{4},\nn
&\ddot{\Delta}^{k}_{(d)eq}=2C_{1}+C_{2}\gamma_{+}(\gamma_{+}-1)\nn
&~~~~~~~+C_{3}\gamma_{-}(\gamma_{-}-1)+\frac{(6-4n)(3-4n)}{9}C_{4}\;.
\end{eqnarray}
Once again, for the sake of simplicity, we do not present them here explicitly.

It turns out that  in the adiabatic  limit, the long wavelength solutions of  the growth factor both in the radiation and dust epochs are exactly the same as those found in \cite{CDT}.
%%%%%%%%%%%%%%%%%%%%%%%%%%%%%%%%%%%%%%%%%%%%%%%%%%%%
\section{Conclusions}
%%%%%%%%%%%%%%%%%%%%%%%%%%%%%%%%%%%%%%%%%%%%%%%%%%%%
In this work we have for the first time presented a detailed analysis of the $(1+3)$- covariant and gauge-invariant theory of cosmological perturbations in situations where the universe is described by a multi-component fluid, with a general equation of state parameter for an arbitrary $f(R)$ theory of gravity. The linearized evolution equations of the density and curvature perturbations of such a universe have been derived for both the fluid components and the total matter, relative to the energy frame. We then have taken the background transient solutions of $R^{n}$ gravity for a two-fluid system dominated respectively by radiation and CDM (dust) and obtained solutions in both the short and long wavelength approximations. These solutions are important when testing a full numerical implementation of these equations,  important for generating the complete matter power spectrum for $f(R)$ gravity theories with a realistic background cosmological expansion history. We also found that for $R^n$ gravity to be consistent with the M\'esz\'aros effect, the parameter $n$ needs to satisfy $n>2/3$.

We also gave a new covariant characterisation of the quasi-static approximation and used this to show that on small scales this approximation is valid for values of $n$ in the neighbourhood of $1$, i.e., it is in good agreement with a numerical integration of the full set of equations for the given set of initial conditions. This is the first time such a quasi-static analysis has been presented in a covariant way both for radiation and dust universes and provided the foundations for detailed comparison with what is found using the metric formalisms,  together with a full computation of the power spectra. This will be presented in a future work.
\ack
The authors thank the National Research Foundation (NRF-South Africa) for financial support. AdlCD also acknowledges Þnancial support from MICINN (Spain) project numbers FPA
2008-00592, FIS 2011-23000 and Consolider-Ingenio MULTIDARK CSD2009-00064.
%%%%%%%%%%%%%%%%%%%%%%%%%%%%%%%%%%%%%%%%%%%%%%%%%%%%
\section*{References}
%%%%%%%%%%%%%%%%%%%%%%%%%%%%%%%%%%%%%%%%%%%%%%%%%%%%

\end{document}